\documentclass[apj]{emulateapj}
\usepackage{apjfonts} 

\newcommand{\n}{\nodata}
\makeindex
\citeindextrue

\bibpunct{(}{)}{,}{a}{}{,}

\shorttitle{Imaging of Quasars and AGN. IV.}
\shortauthors{Kovalev et al.}

\submitted{}
\received{March 23, 2005}
\accepted{July 7, 2005}
\journalinfo{Astronomical~journal}

\defcitealias{Kellermann_etal98}{Paper~I}
\defcitealias{Zensus_etal02}{Paper~II}
\defcitealias{Kellermann_etal04}{Paper~III}

\begin{document}

\title{Sub-milliarcsecond Imaging of Quasars and Active Galactic Nuclei\\
IV. Fine Scale Structure}


\author{Y. Y. Kovalev\altaffilmark{1,2},
        K. I. Kellermann\altaffilmark{3},
        M. L. Lister\altaffilmark{4},
        D. C. Homan\altaffilmark{5},
        R. C. Vermeulen\altaffilmark{6},
        M. H. Cohen\altaffilmark{7},
	E. Ros\altaffilmark{8},
        M. Kadler\altaffilmark{8},
	A. P. Lobanov\altaffilmark{8},
        J. A. Zensus\altaffilmark{8,3},
        N. S. Kardashev\altaffilmark{2},
        L. I. Gurvits\altaffilmark{9},
        M. F. Aller\altaffilmark{10},
        and
	H. D. Aller\altaffilmark{10}
	}

\altaffiltext{1}{Jansky Fellow,
                 National Radio Astronomy Observatory,
                 P.O.~Box 2, Green Bank, WV 24944, U.S.A.;
		 \mbox{ykovalev@nrao.edu}
		 }

\altaffiltext{2}{Astro Space Center of Lebedev Physical Institute,
                 Profsoyuznaya 84/32, 117997 Moscow, Russia;
		 \mbox{nkardash@asc.rssi.ru}
		 }

\altaffiltext{3}{National Radio Astronomy Observatory,
                 520 Edgemont Road, Charlottesville, VA~22903--2475, U.S.A.;
                 \mbox{kkellerm@nrao.edu}
		 }

\altaffiltext{4}{Department of Physics, Purdue University,
                 525 Northwestern Avenue, West Lafayette, IN 47907, U.S.A.;
                 \mbox{mlister@physics.purdue.edu}
		 }

\altaffiltext{5}{Department of Physics and Astronomy, Denison University,
                 Granville, OH 43023, U.S.A.;
                 \mbox{homand@denison.edu}
		 }

\altaffiltext{6}{ASTRON,
                 Netherlands Foundation for Research in Astronomy,
                 P.O.~Box 2, NL-7990 AA Dwingeloo, The Netherlands;
                 \mbox{rvermeulen@astron.nl}
		 }

\altaffiltext{7}{Department of Astronomy, Mail Stop 105-24,
                 California Institute of Technology,
                 Pasadena, CA 91125, U.S.A.;
                 \mbox{mhc@astro.caltech.edu}
		 }

\altaffiltext{8}{Max-Planck-Institut f\"ur Radioastronomie, Auf dem H\"ugel 69,
                 D-53121 Bonn, Germany;
                 \mbox{eros@mpifr-bonn.mpg.de},
		 \mbox{mkadler@mpifr-bonn.mpg.de},
		 \mbox{alobanov@mpifr-bonn.mpg.de},
                 \mbox{azensus@mpifr-bonn.mpg.de}
		 }

\altaffiltext{9}{Joint Institute for VLBI in Europe, P.O.~Box 2,
                 7990 AA Dwingeloo, The Netherlands;
                 \mbox{lgurvits@jive.nl}
		 }

\altaffiltext{10}{Department of Astronomy, University of Michigan,
                  830 Dennison Building, Ann Arbor, MI 48109--1090, U.S.A.;
                  \mbox{mfa@umich.edu}, \mbox{haller@umich.edu}
		  }

\begin{abstract}

     We have examined the compact structure in 250 flat-spectrum
extragalactic radio sources using interferometric fringe visibilities
obtained with the VLBA at 15~GHz. With projected baselines out to 440
million wavelengths, we are able to investigate source structure on
typical angular scales as small as 0.05~mas. This scale is similar to
the resolution of VSOP space VLBI data obtained on longer baselines
at a lower frequency and with somewhat poorer accuracy. For 171 sources
in our sample, more than half of the total flux density seen by the
VLBA remains unresolved on the longest baselines. There are 163 sources
in our list with a median correlated flux density at 15~GHz in excess
of 0.5~Jy on the longest baselines; these will be useful as
fringe-finders for short wavelength VLBA observations. The total flux
densities recovered in the VLBA images at 15~GHz are generally close to
the values measured around the same epoch at the same frequency with the
\mbox{RATAN--600} and UMRAO radio telescopes.

     We have modeled the core of each source with an elliptical
Gaussian component. For about 60\%\ of the sources, we have at least
one observation in which the core component appears unresolved
(generally smaller than 0.05~mas) in one direction, usually transverse
to the direction into which the jet extends. BL~Lac objects are on
average more compact than quasars, while active galaxies are on average
less compact. Also, in an active galaxy the sub-milliarcsecond core
component tends to be less dominant. Intra-Day Variable (IDV) sources
typically have a more compact, more core-dominated structure on
sub-milliarcsecond scales than non-IDV sources, and sources with a
greater amplitude of intra-day variations tend to have a greater
unresolved VLBA flux density. The objects known to be GeV gamma-ray loud
appear to have a more compact VLBA structure than the other sources in
our sample. This suggests that the mechanisms for the production of
gamma-ray emission and for the generation of compact radio synchrotron
emitting features are related.

     The brightness temperature estimates and lower limits for the
cores in our sample typically range between $10^{11}$ and $10^{13}$\,K,
but they extend up to $5\,\times\,10^{13}$\,K, apparently in excess of
the equipartition brightness temperature, or the inverse Compton limit
for stationary synchrotron sources. The largest component speeds are
observed in radio sources with high observed brightness temperatures,
as would be expected from relativistic beaming. Longer baselines, which
may be obtained by space VLBI observations, will be needed to resolve
the most compact high brightness temperature regions in these sources.

\end{abstract}

\section{Introduction}

\begin{figure}[t]
\begin{center}
\resizebox{1.0\hsize}{!}{
\includegraphics[trim=0cm -1cm 0cm 0cm]{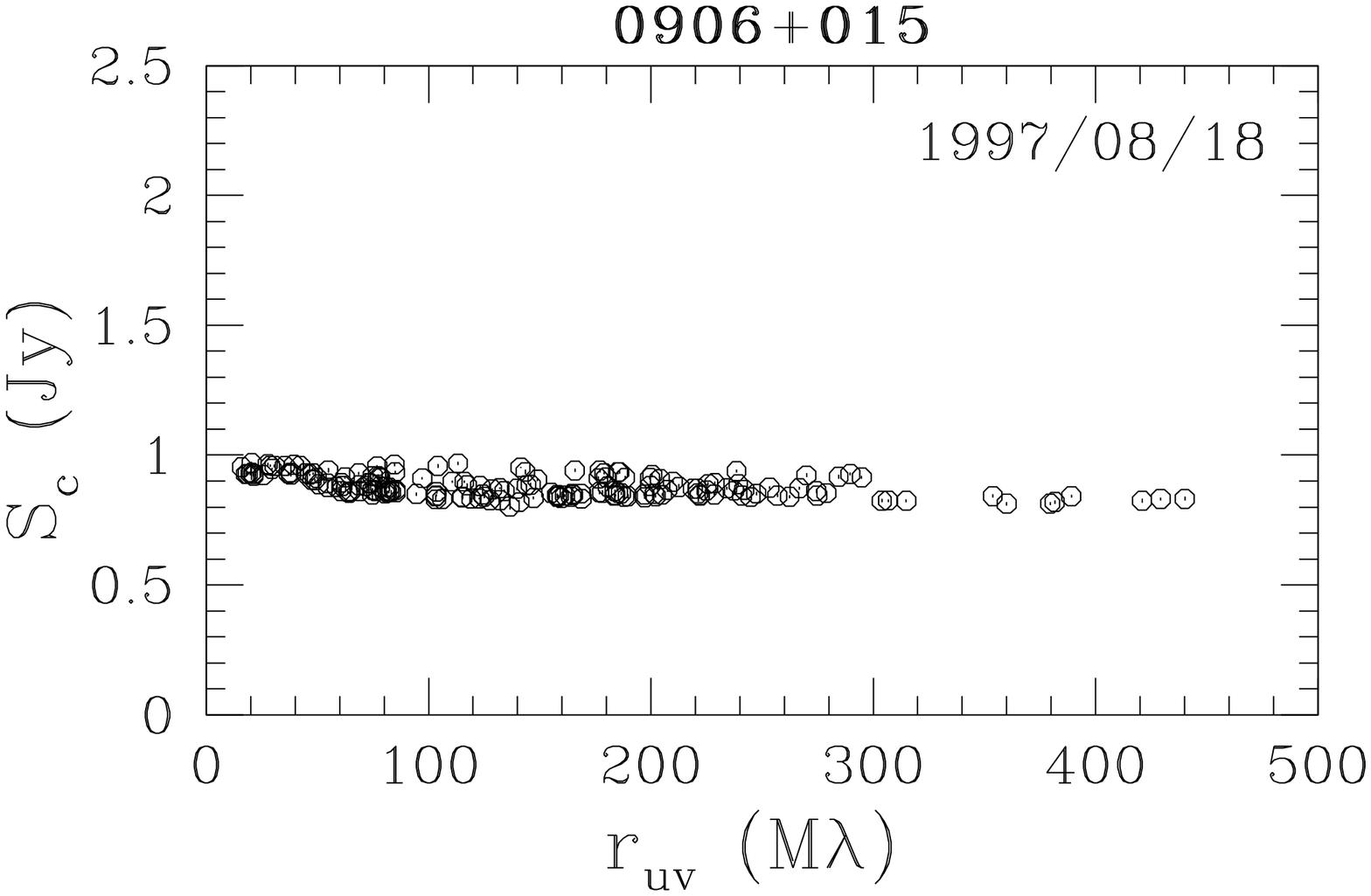}
\includegraphics[trim=0cm -1cm 0cm 0cm]{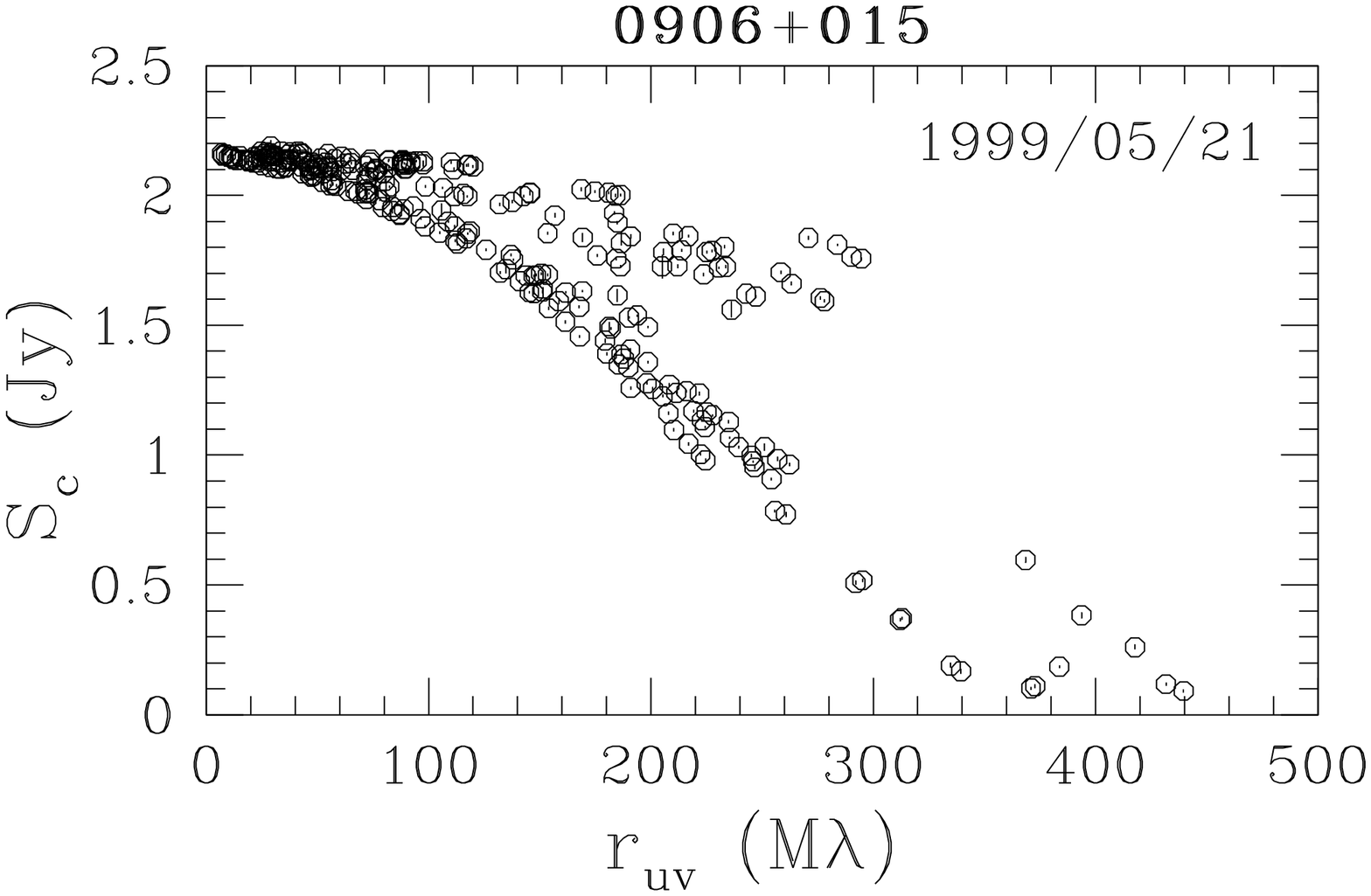}
}
\resizebox{1.0\hsize}{!}{
\includegraphics[trim=-0.2cm 0cm 0cm 0cm]{f1c.eps}
\includegraphics[trim=-0.2cm 0cm 0cm 0cm]{f1d.eps}
}
\end{center}
\caption{
Comparison of two epochs for the quasar 0906+015. Top: visibility
function amplitude (i.e.\ correlated flux density), $S_\mathrm{c}$,
versus projected spacing, $r_{uv}$.
Bottom: naturally weighted CLEAN images.
     The lowest contour is plotted at the level of 0.16\% and 0.13\% of
the peak brightness of 0.87 (epoch 1997/08/18) and 1.79 (epoch
1999/05/21) Jy/beam, respectively. Other contours are shown at
increasing powers of $\sqrt{2}$. The half-power width of the
synthesized beam (with natural weighting) is shown in the left lower corners.
     The ($u$,$v$)-coverage of the two experiments was similar.
     One can clearly see (and model) from the $S_\mathrm{c}$--$r_{uv}$
plots that an unresolved component observed at the first epoch at a
level of about 0.8~Jy is heavily resolved at the second epoch, which is
not clear from the corresponding CLEAN images. This component dominates
the $S_\mathrm{c}$--$r_{uv}$ dependence.
}
\label{radplot-map}
\end{figure}

     Early interferometric observations of radio source structure were
typically analyzed by examining how the amplitude of the fringe
visibility varied with projected interferometer spacing
\citep[e.g.,][]{Rowson63}. Although these techniques for conventional
connected interferometers were later replaced by full synthesis imaging
incorporating Fourier inversion, CLEAN \citep[e.g.,][]{Hogbom74}, and
self calibration \citep[see review by][]{PearsonRedhead84}, the
interpretation of the early VLBI observations, again, was based on the
examination of fringe amplitudes alone
\citep[see][]{Cohen_etal75}. Indeed, the discovery of superluminal
motion in the source \objectname{3C\,279} \citep{Cohen_etal71,
Whitney_etal71} was based on single baseline observations of the change
in spacing of the first minimum of the fringe visibility. However,
after the development of phase-closure techniques, reliable full
synthesis images have been produced from VLBI observations for more
than 25 years. However, these tend to hide the information on the
smallest scale structures, because of the convolution with the
synthesized beam (e.g., Figure~\ref{radplot-map}). A more thorough
discussion of non-imaging VLBI data analysis is given by
\cite{Pearson99}.

     The best possible angular resolution is needed to study the
environment close to supermassive black holes where relativistic
particles are accelerated and collimated to produce radio jets. The
greatest angular resolution to date was obtained in observations of
interstellar scintillations \citep[e.g.,][]{KC_etal97,Macquart_etal00,
KC_etal01, JM01, Rickett_etal01, DB02, Kraus_etal03, Lovell_etal03}.
The resolution achievable with VLBI can be improved by observing at
shorter wavelengths \citep[e.g.,][]{Moellenbrock_etal96,
Lobanov_etal00,Lister01,Greve_etal02} or by increasing the physical
baseline lengths using Earth-to-space interferometry. The first space
VLBI missions \citep{Levy_etal89,Hirabayashi_etal98} increased the
available baseline lengths by a factor of about 3. Planned space VLBI
observations such as RadioAstron \citep{RadioAstron}, VSOP--2
\citep{VSOP2}, and ARISE \citep{ARISE}, will extend the baselines
further.

     For simple source structures, a direct study of the fringe
visibilities can give a better angular resolution than an analysis of
the images reconstructed from these data
\citep[e.g.,][]{MaltbyMoffet62}. In principle, a careful deconvolution
of the images should give equivalent results. However, experience has
shown that when confronted with even moderately complex images, it is
dangerous to attempt to increase the resolution significantly beyond
that of the CLEAN restoring beam size; a procedure referred to in early
radio astronomy literature as ``super resolution''.

     In two previous papers (\citealt*{Kellermann_etal98}, hereafter
\citetalias{Kellermann_etal98}, and \citealt*{Zensus_etal02}, hereafter
\citetalias{Zensus_etal02}) we have described the
sub-milliarcsecond scale structure of 171 active galactic nuclei, based
on naturally weighted images made from observations with the VLBA
\citep{VLBA} at 15~GHz. In addition, in \citetalias{Zensus_etal02} we
have placed more restrictive limits on the sizes of unresolved
sources by direct analysis of the fringe visibilities. In a third paper
(\citealt*{Kellermann_etal04}, hereafter
\citetalias{Kellermann_etal04}) we have reported on the observed
motions in the jets of 110 of these sources during the period 1994 to
2001.

     In this paper, we analyze 15~GHz VLBA observations of the central
regions of 250 extragalactic radio sources.
We use the visibility function
data to study the most compact structures and the way they change with
time. The smallest features we are able to discern from these data have
an extent of about 0.02--0.06~mas. For the nearest object in our study,
\objectname{1228+126} (\objectname{M\,87}, \objectname{Virgo\,A}), this
corresponds to a linear size of $10^{16}$~cm, or several tens of
Schwarzschild radii, if the mass of the central object is
$3\,\times\,10^9$ solar masses and the distance is 17.5~Mpc. We define
our sample in \S~\ref{sampledef}, describe the visibility data in
\S~\ref{visdata}, and the model fitting and analysis in
\S~\ref{modelfitting}. In \S~\ref{results} we discuss the results, and
the conclusions are summarized in \S~\ref{summary}.

     Throughout this paper we use the following cosmological
parameters: $H_0 = 70$~km\,s$^{-1}$\,Mpc$^{-1}$,
$\Omega_\mathrm{m}=0.3$, and $\Omega_\Lambda=0.7$.  We adopt the
convention of using the term ``quasar'' to describe optical
counterparts brighter than absolute magnitude $-23$, and ``active
galaxy'' for the fainter objects.

\section{Sample Definition}
\label{sampledef}

     Our analysis is based on data obtained during the period
1994--2003 as part of the VLBA 15~GHz monitoring survey of
extragalactic sources (Papers I, II, and III, \citealt*{MOJAVE-I},
E.~Ros et al., in preparation). We have also used additional
observations made in 1998 and 1999 by L.~I.\ Gurvits et al.~(in
preparation) as part of a separate program to compare 15~GHz source
structure measured with the VLBA to 5~GHz structure measured in the
framework of the VSOP Survey Program
\citep{Hirabayashi_etal00,Lovell_etal2004,Scott_etal2004,Horiuchi_etal2004}.
The program by Gurvits et al.\ used the same observing and data
reduction procedures as the VLBA 15~GHz monitoring survey, and provides
both additional sources and additional epochs.

     Our dataset consists of 1204 VLBA observations of 250 different
compact extragalactic radio sources. The initial calibration of the
data was carried out with the NRAO $\cal{AIPS}$ package \citep{aips},
and was followed by imaging with the DIFMAP program~\citep{difmap},
mostly with the use of an automatic script
\citepalias{Kellermann_etal98,Zensus_etal02}. The CLEAN images as well
as the visibility function data are available on our web
sites\footnote{{\sf{}\url{http://www.nrao.edu/2cmsurvey/}} and
{\sf{}\url{http://www.physics.purdue.edu/astro/MOJAVE/}}}.

     Most of the radio sources contained in our ``full sample'' of 250
sources have flat radio spectra ($\alpha>-0.5$, $S_\nu\sim\nu^\alpha$),
and a total flux density at 15~GHz (often originally estimated by
extrapolation from lower frequency data) greater than 1.5~Jy for
sources with declination $\delta>0^\circ$, or greater than 2~Jy for
sources with $-20^\circ<\delta<0^\circ$. However, additional sources
which did not meet these criteria but are of special interest were also
included in the full sample.

     Our full sample is useful for investigating fine scale structure
in a cross-section of known extragalactic radio source classes, and for
planning future (space) VLBI observations. However, in order to compare
observations with the theoretical predictions of relativistic beaming
models, it is also useful to have a well-defined sub-sample selected on
the basis of beamed, rather than total, flux density. We have therefore
formed a flux density limited complete sample which has been used as
the basis of our jet monitoring program since mid 2002, called ``The
MOJAVE Program: {\bf{}M}onitoring {\bf{}O}f {\bf{}J}ets in {\bf{}A}GN
with {\bf{}V}LBA {\bf{}E}xperiments'' (\citetalias{Kellermann_etal04},
\citealt*{MOJAVE-I}).  There are 133 sources in the MOJAVE sub-sample.
The redshift distribution for these sources ranges up to 3.4 (quasar
\objectname{0642+449}), although most sources have redshifts less than
2.5, with a peak in the distribution near 0.8.

     Table~\ref{sample} summarizes the properties of each source.
Columns 1 and 2 give the IAU source designation, and where appropriate,
a commonly used alias; J2000.0 coordinates are in columns 3 and 4. The
optical classification and redshift are shown in columns 5 and 6,
respectively; these were obtained mainly from \cite{VCV03}, as
discussed below. In column 7 we give a radio spectral classification
for each source based on the \mbox{RATAN--600} radio telescope
observations of broad-band instantaneous spectra from 1 to 22~GHz
\citep{Kovalev_etal99,Kovalev_etal2000}. These spectra are available on
our web site. For the few sources which were not observed at
\mbox{RATAN--600}, we used published (non-simultaneous) radio flux
densities taken from the literature. We consider a radio spectrum
to be ``flat'' if any portion of its spectrum in the range 0.6~GHz to
22~GHz has a spectral index flatter than $-0.5$ and ``steep'' if the
radio spectral index is steeper than $-0.5$ over this entire region. In
column 8 we indicate whether or not the radio source is associated with
a gamma-ray detection by EGRET \citep{Mattox_etal01,SRM03,SRM04}.
Columns 9 and 10 indicate whether or not the source is a member of the
complete correlated flux density limited MOJAVE sample and the VSOP
5~GHz AGN survey source sample
\citep{Hirabayashi_etal00,Lovell_etal2004,Scott_etal2004,Horiuchi_etal2004}.
Column 11 gives references to papers reporting intra-day variability
(IDV) of the source total flux density.

     Of the 250 sources in the full sample, there are 179 quasars, 37
BL~Lacertae objects, 23 active galaxies, and 11 sources which are
optically unidentified. The MOJAVE complete sample of 133 sources
includes 94 quasars, 22 BL~Lacertae objects, 8 active galaxies, and 9
unidentified objects. These classifications come from \cite{VCV03}, who
defined a quasar as a star-like object, or an object with a star-like
nucleus, with broad emission lines, brighter than absolute magnitude
$\mathrm{M_B}=-23$.

     \cite{VCV03} provide a list of BL~Lacertae objects, which
historically were defined as bright galactic nuclei which are highly
polarized in the optical regime, and for which no emission or
absorption lines have been detected. The precise delineation between
BL~Lacs and OVV quasars remains controversial \citep{VCV00}, since the
original proposed 5\,\AA\ limit \citep{SFK91} is arbitrary
\citep{SF97}, and individual emission line equivalent widths are now
known to be highly variable over time. Indeed the prototype, BL~Lac
itself, shows broad and narrow emission lines, as well as stellar
absorption lines, in modern spectra: it no longer meets the classical
definition of a BL~Lac \citep{Vermeulen_etal95}. The detectability of
narrow and broad emission lines and absorption lines is set to a very
significant degree by the variable continuum level, signal-to-noise
ratio (SNR), starlight contribution, and other extrinsic and
time-dependent factors \citep[see, e.g.,][]{MB95}. The situation is
complicated further by proposed unification schemes
\citep{UrryPadovani95}, which apparently led \cite{VCV03} to
re-classify many BL Lacs as quasars solely on the basis of their
extended 5~GHz luminosity being above the FR~I/II division. 

     Many of the objects in our sample are blazars, which are defined
as the union of the original categories of BL~Lacertae objects and
optically violently variable (OVV) quasars. Both groups are highly
polarized and variable in the optical spectral region
\citep[e.g.,][]{AS80}. Since we are interested in comparing the radio
properties of strong and weak-lined blazars, we retain the original
BL~Lac classifications for these objects and indicate these and other
controversial classifications in the notes to Table~\ref{sample}. For
our analysis, it would have been preferable to directly use optical
line equivalent width data; however, high-quality, multi-epoch spectra
are currently available for only a small fraction of our sample.
Nevertheless, the objects originally classified as BL Lacs do, on
average, have lower equivalent width spectral lines than classical
quasars. This seems to be (i) partly the effect of dilution by a beamed
non-thermal continuum \citep[e.g.,][]{Wills_etal83}, and (ii) partly
because many of these objects are intrinsically different from classical
quasars as shown by their diffuse radio emission which is similar to
that of FR~I radio galaxy \citep[e.g.,][]{Kollgaard_etal92,RS2001}.
These two effects cannot be clearly separated using only VLBI data. We
show here that on average objects historically called ``BL Lacs''
differ statistically from classical quasars in their parsec scale radio
properties. Physical interpretation depends on separating the above two
effects.

\section{Visibility functions}
\label{visdata}

\begin{figure*}[p]
\resizebox{1.04\hsize}{!}{\includegraphics[trim=0cm 1.0cm 0cm 1.5cm]{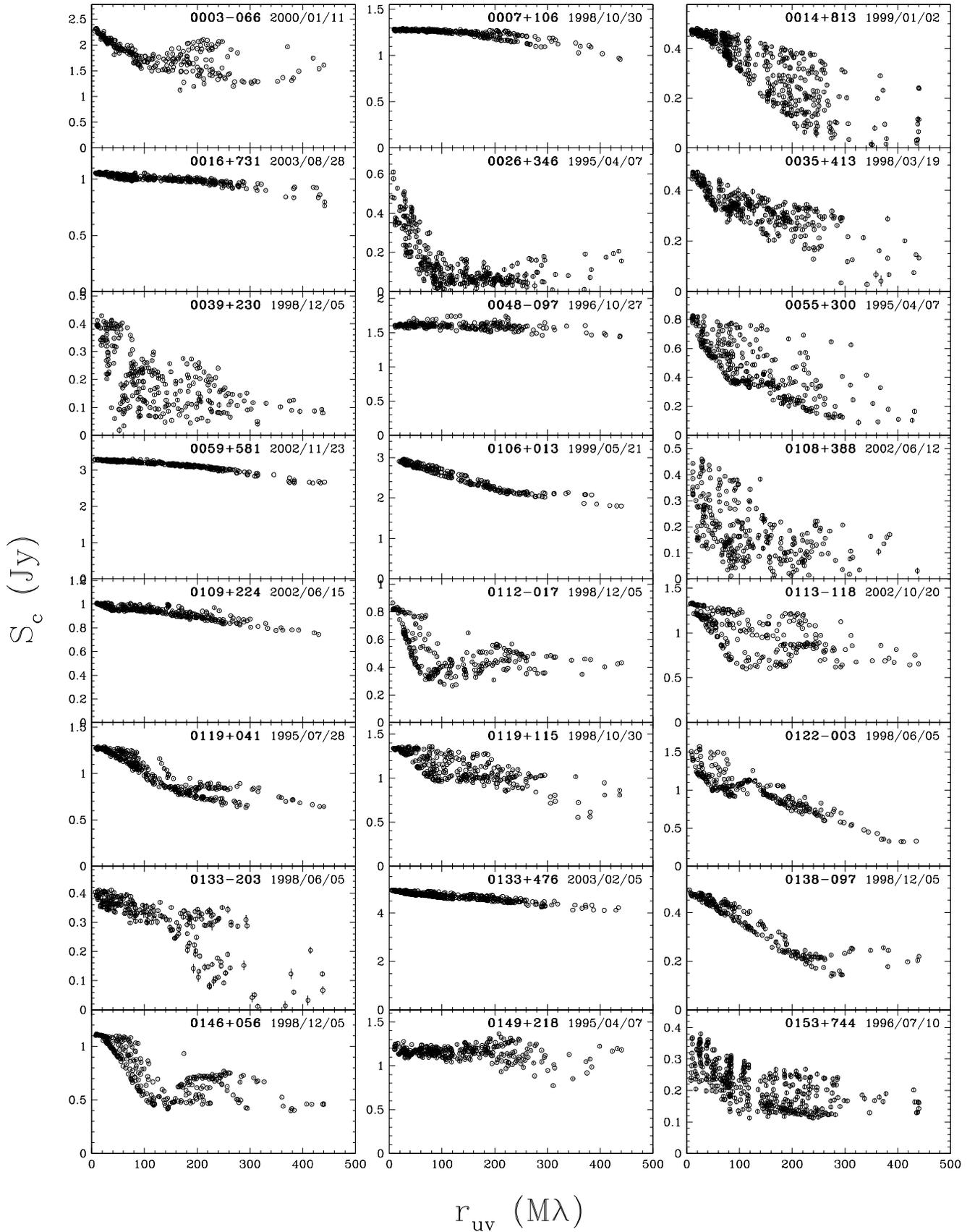}}
\caption{
\footnotesize
Amplitude of the visibility function (i.e.\ correlated flux
density), $S_\mathrm{c}$, versus projected spacing, $r_{uv}$. Each point
represents a coherent average over one 4--6 min observation on an
individual interferometer baseline. The error bars, which represent
only the statistical errors, are often smaller than the symbol. Errors
of the absolute flux density calibration are not shown here and are
about 5\%. For each source, the data are presented at the epoch when
$S_\mathrm{unres}$ is maximum. The plots for all the epochs observed in
the 15~GHz VLBA monitoring program until 2003/08/28 are shown in the
electronic version.
}
\label{Auv}
\end{figure*}

\begin{figure*}[p]
\figurenum{\arabic{figure}}
\resizebox{1.04\hsize}{!}{\includegraphics[trim=0cm 1cm 0cm 1.5cm]{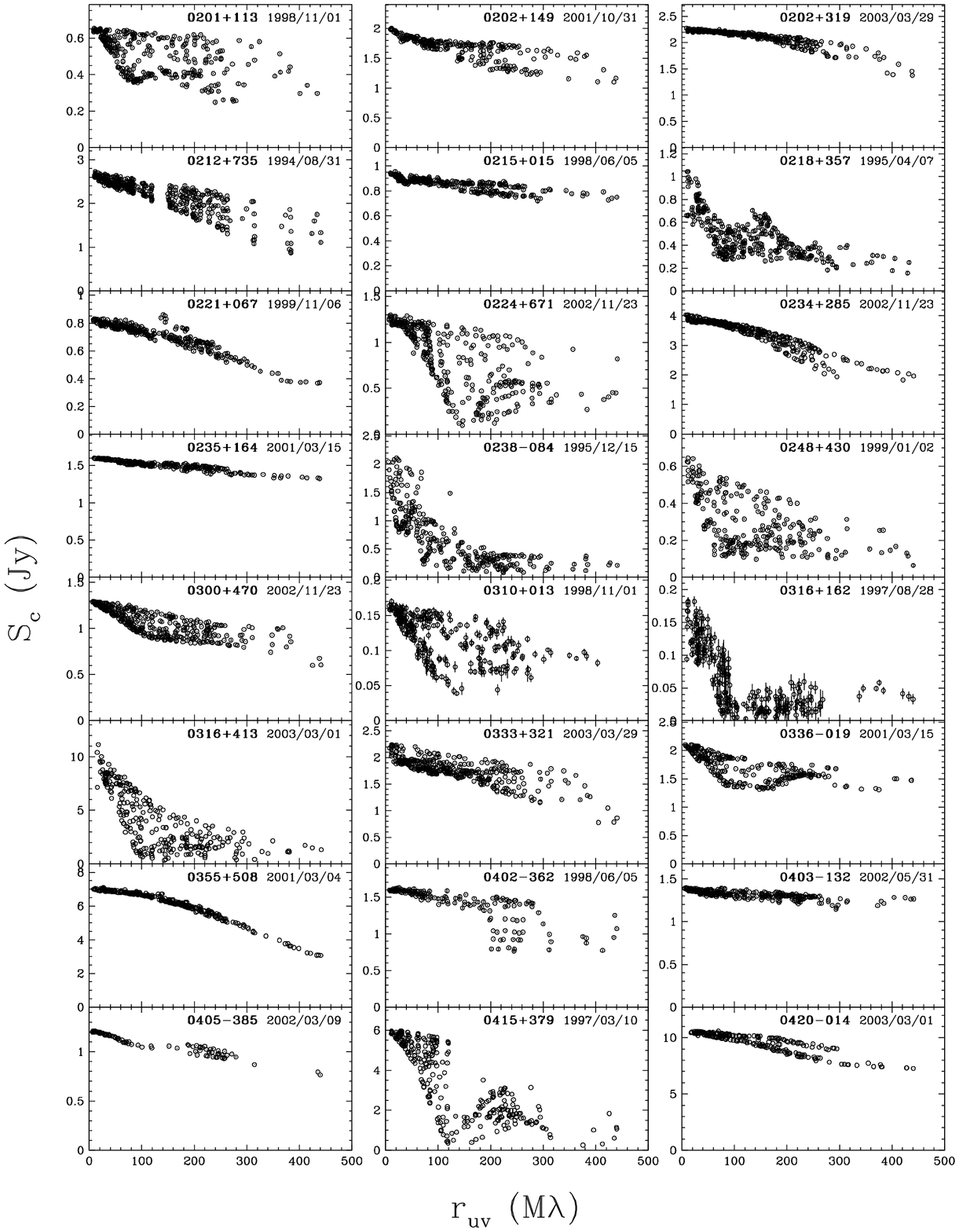}}
\caption{Continued}
\end{figure*}

\figurenum{\arabic{figure}}
\begin{figure*}[p]
\resizebox{1.04\hsize}{!}{\includegraphics[trim=0cm 1cm 0cm 1.5cm]{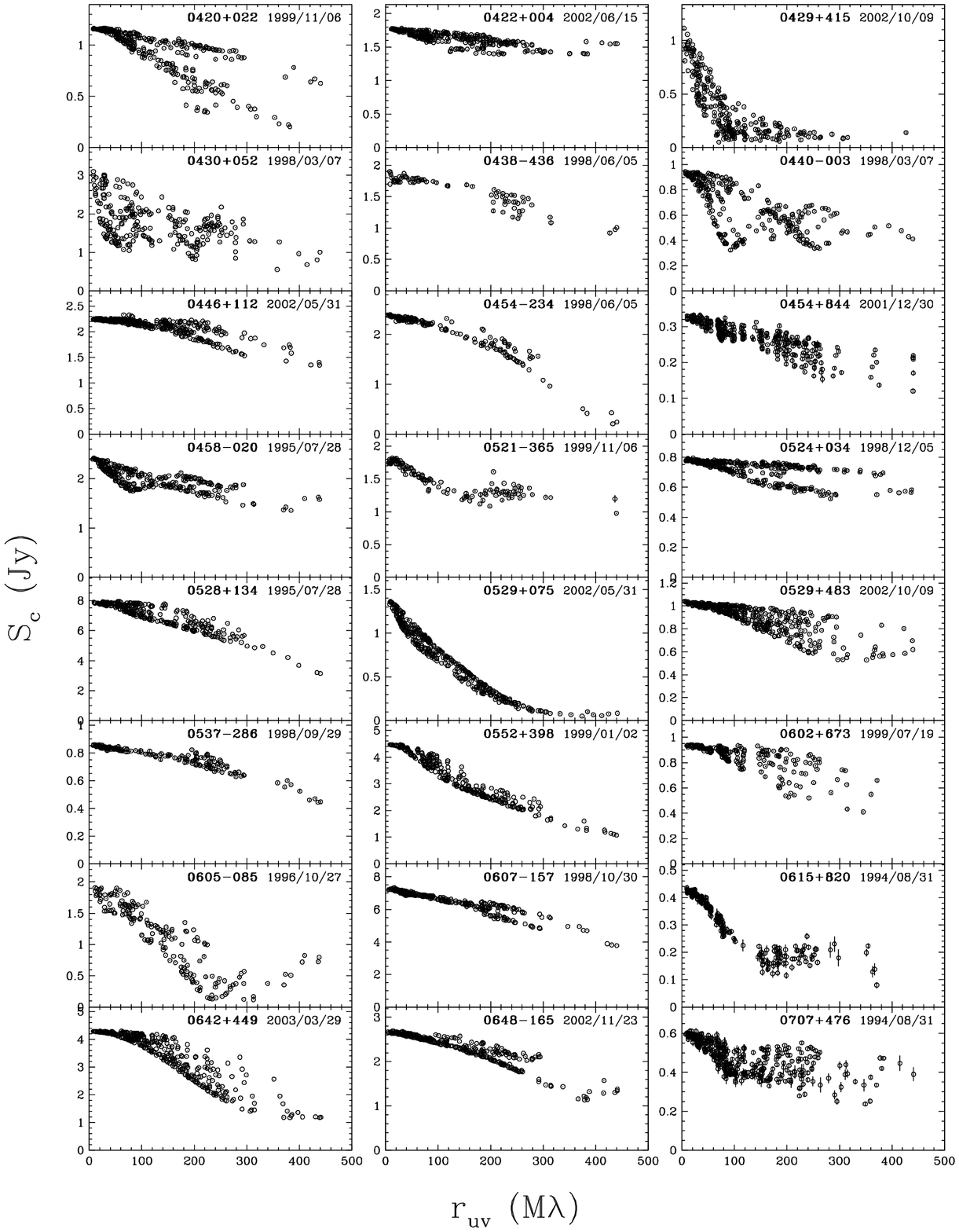}}
\caption{Continued}
\end{figure*}

\figurenum{\arabic{figure}}
\begin{figure*}[p]
\resizebox{1.04\hsize}{!}{\includegraphics[trim=0cm 1cm 0cm 1.5cm]{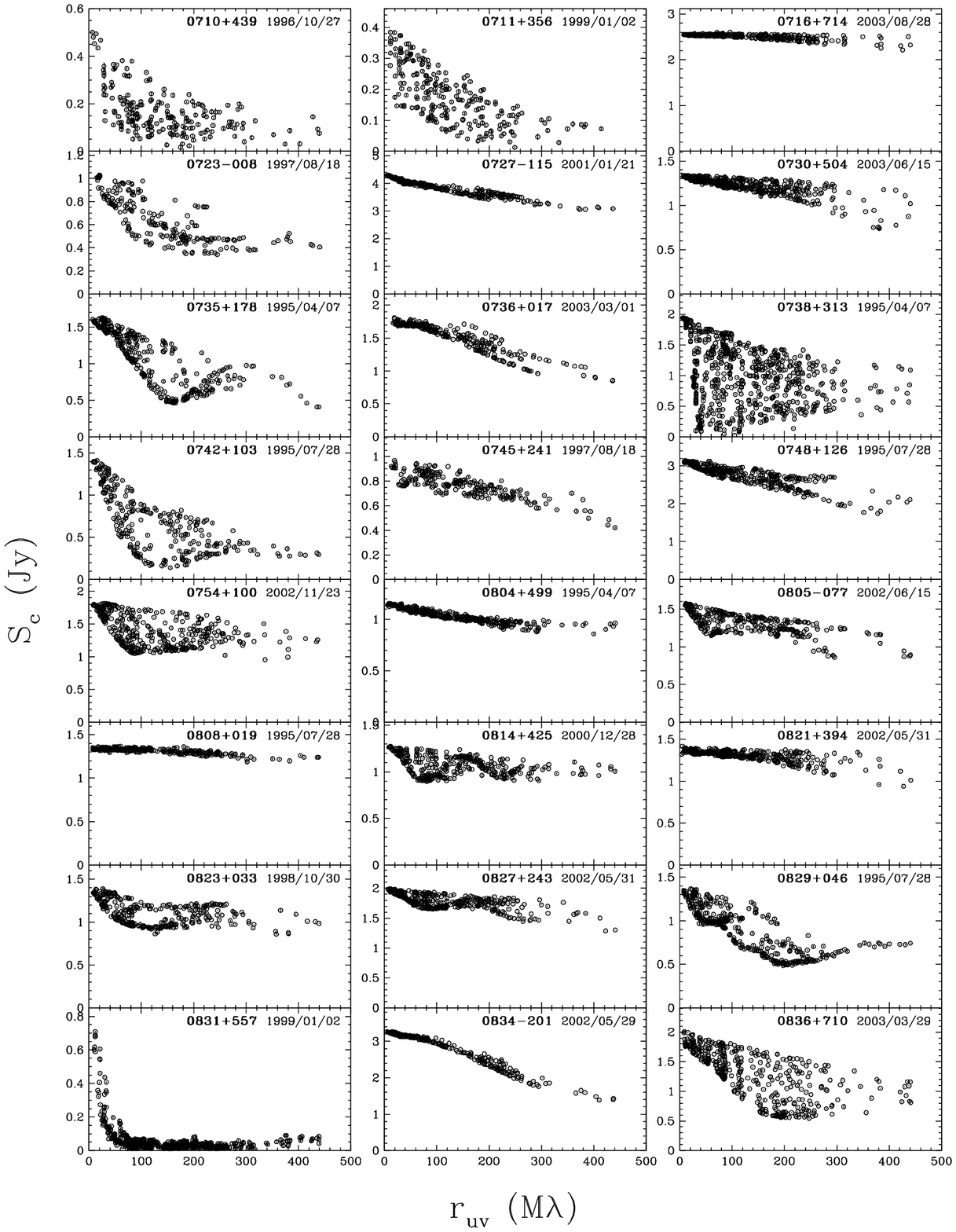}}
\caption{Continued}
\end{figure*}

\figurenum{\arabic{figure}}
\begin{figure*}[p]
\resizebox{1.04\hsize}{!}{\includegraphics[trim=0cm 1cm 0cm 1.5cm]{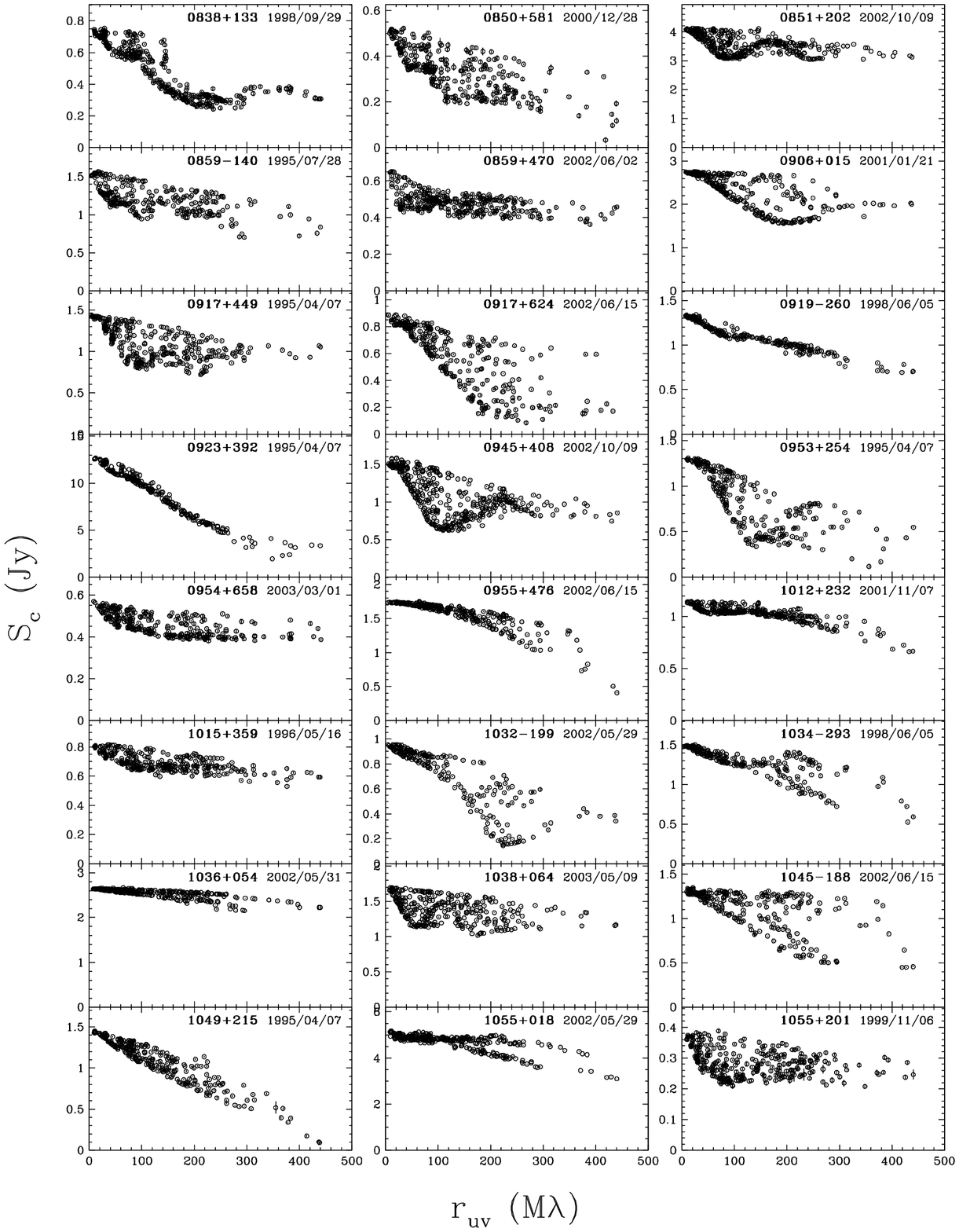}}
\caption{Continued}
\end{figure*}

\figurenum{\arabic{figure}}
\begin{figure*}[p]
\resizebox{1.04\hsize}{!}{\includegraphics[trim=0cm 1cm 0cm 1.5cm]{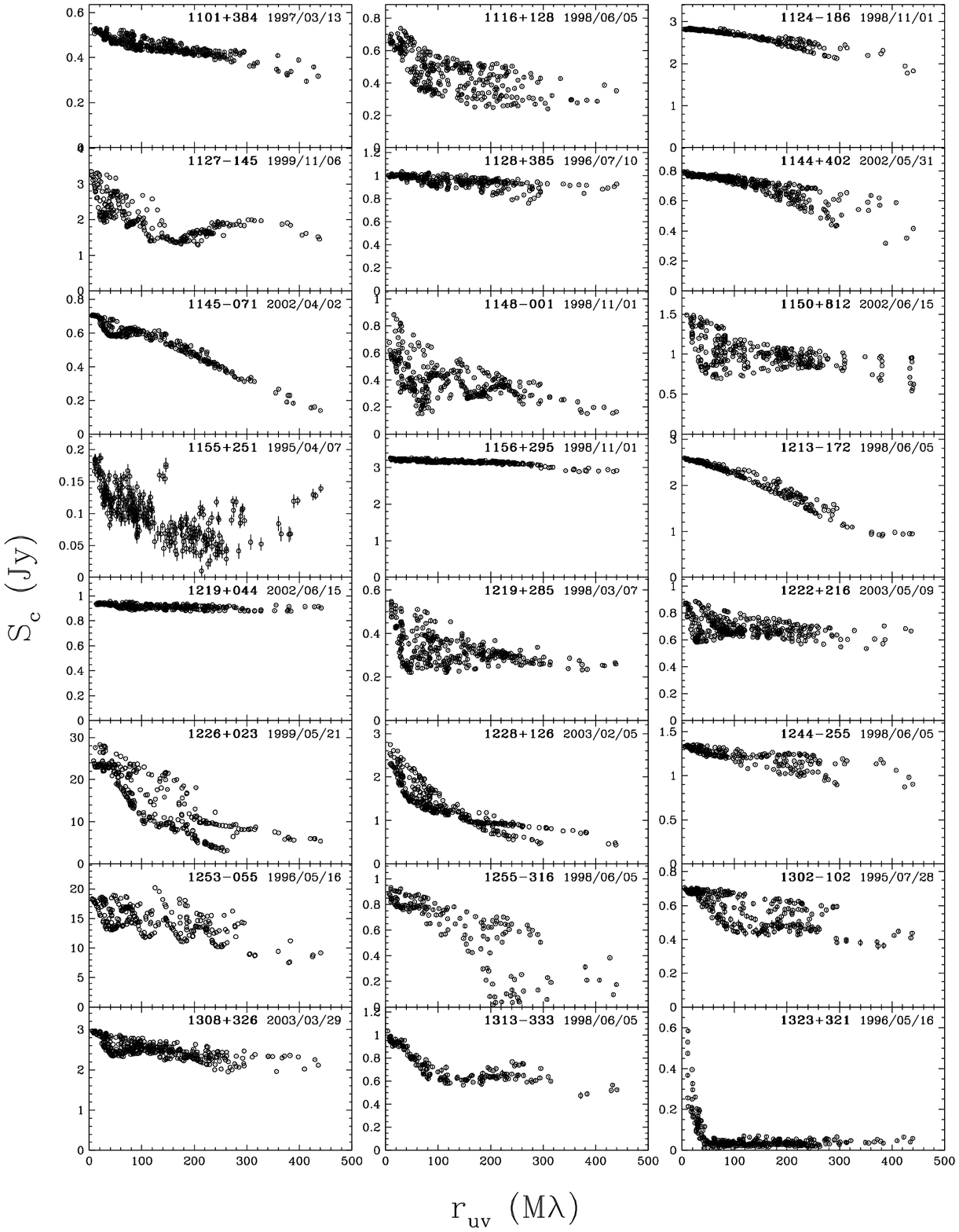}}
\caption{Continued}
\end{figure*}

\figurenum{\arabic{figure}}
\begin{figure*}[p]
\resizebox{1.04\hsize}{!}{\includegraphics[trim=0cm 1cm 0cm 1.5cm]{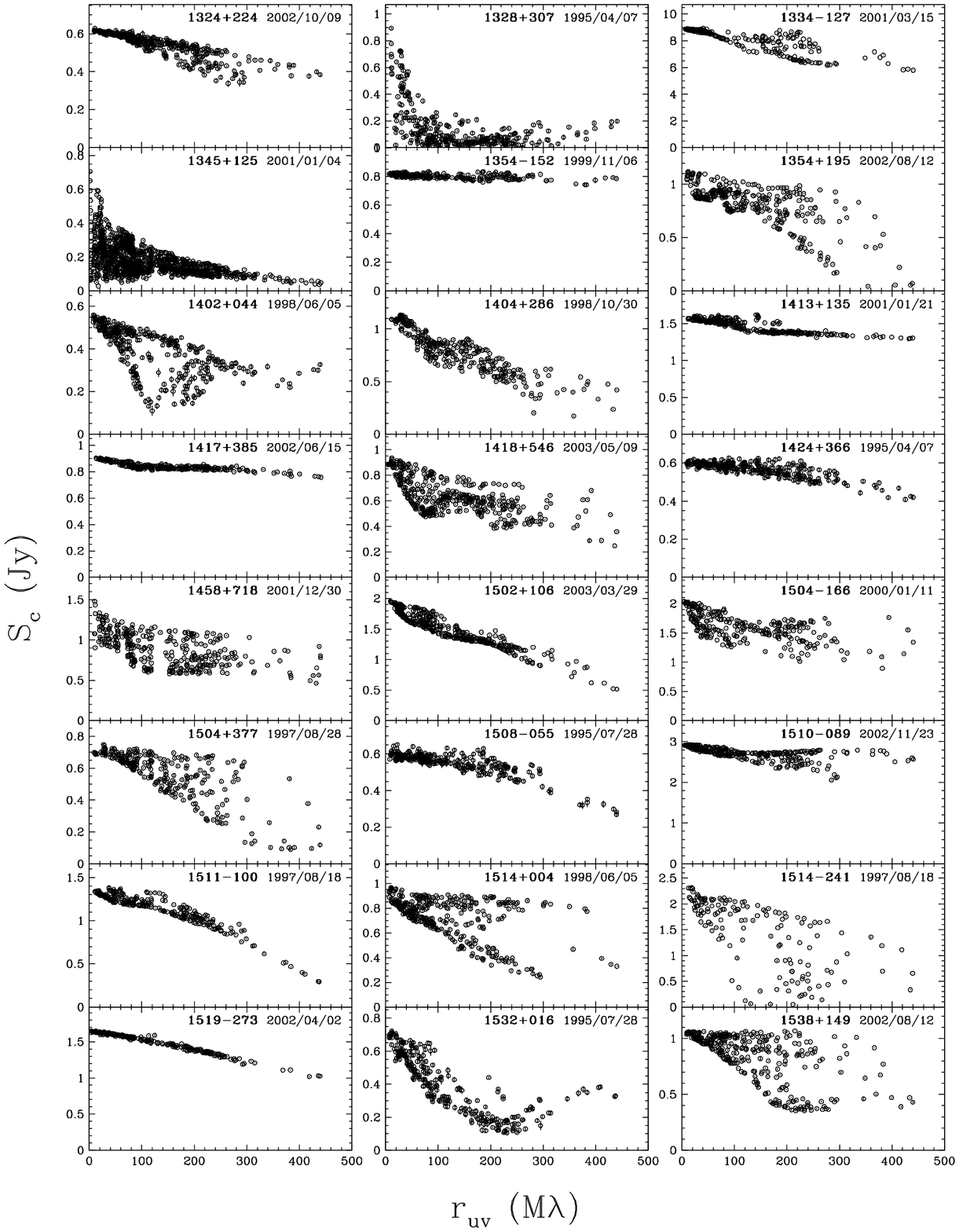}}
\caption{Continued}
\end{figure*}

\figurenum{\arabic{figure}}
\begin{figure*}[p]
\resizebox{1.04\hsize}{!}{\includegraphics[trim=0cm 1cm 0cm 1.5cm]{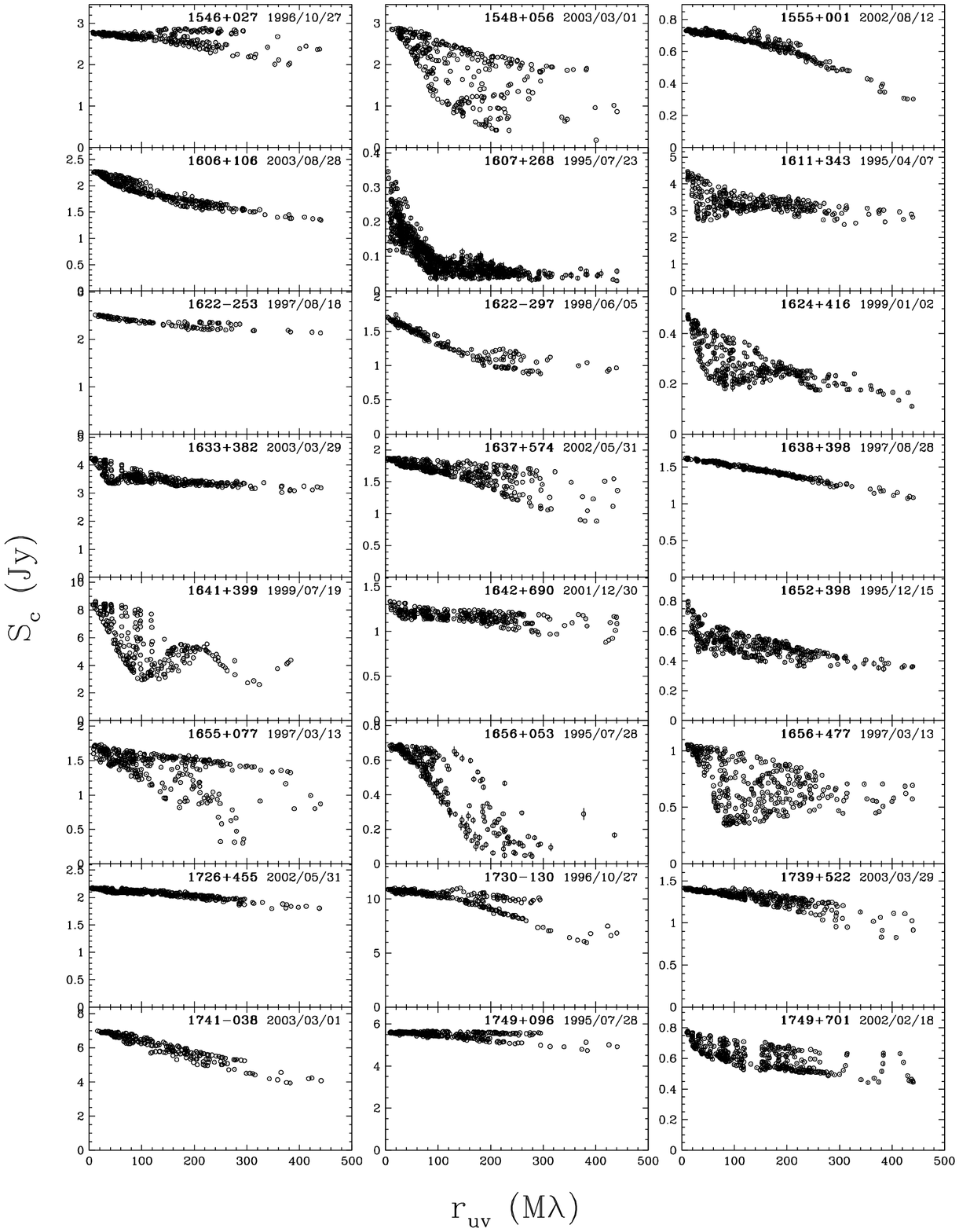}}
\caption{Continued}
\end{figure*}

\figurenum{\arabic{figure}}
\begin{figure*}[p]
\resizebox{1.04\hsize}{!}{\includegraphics[trim=0cm 1cm 0cm 1.5cm]{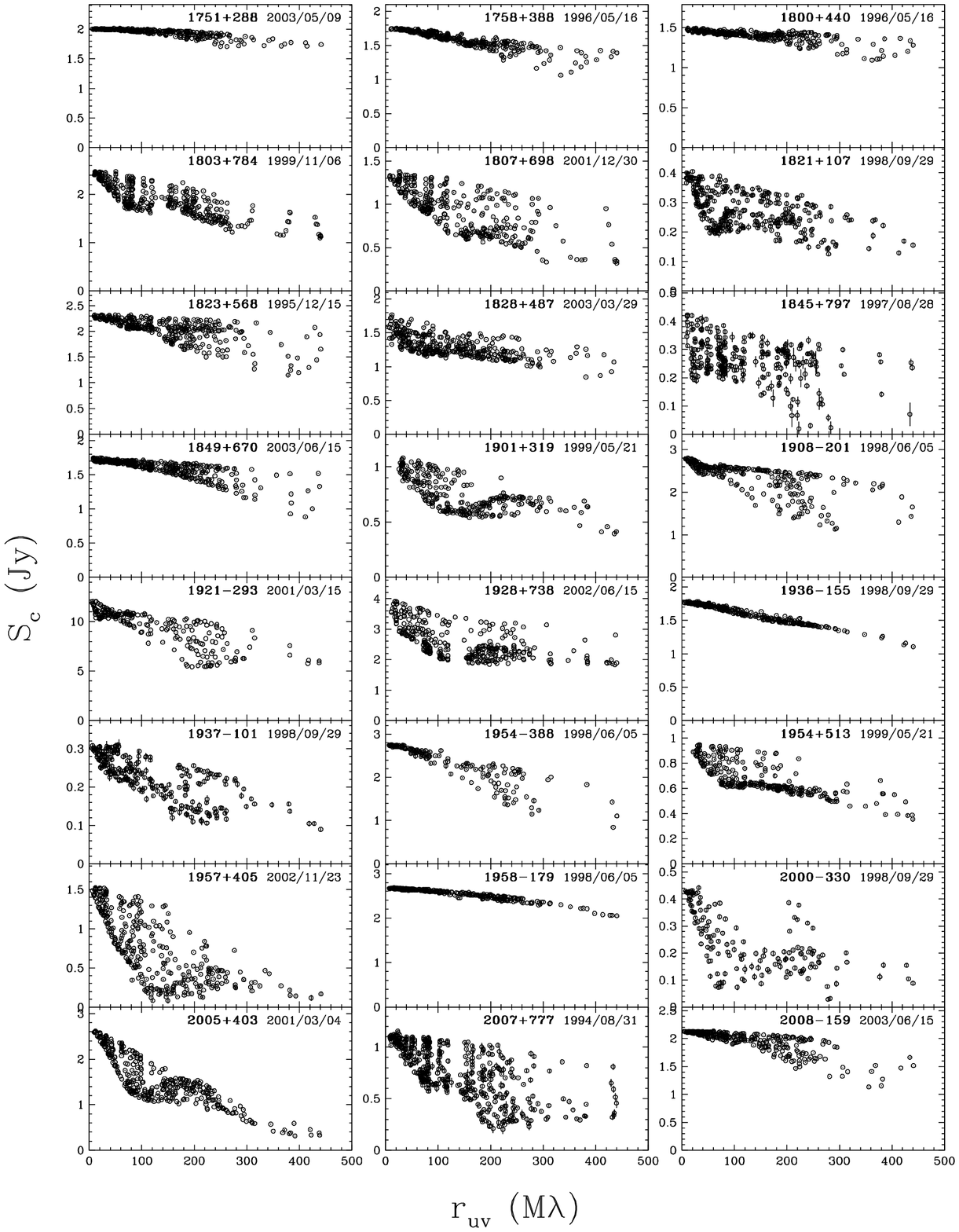}}
\caption{Continued}
\end{figure*}

\figurenum{\arabic{figure}}
\begin{figure*}[p]
\resizebox{1.04\hsize}{!}{\includegraphics[trim=0cm 1cm 0cm 1.5cm]{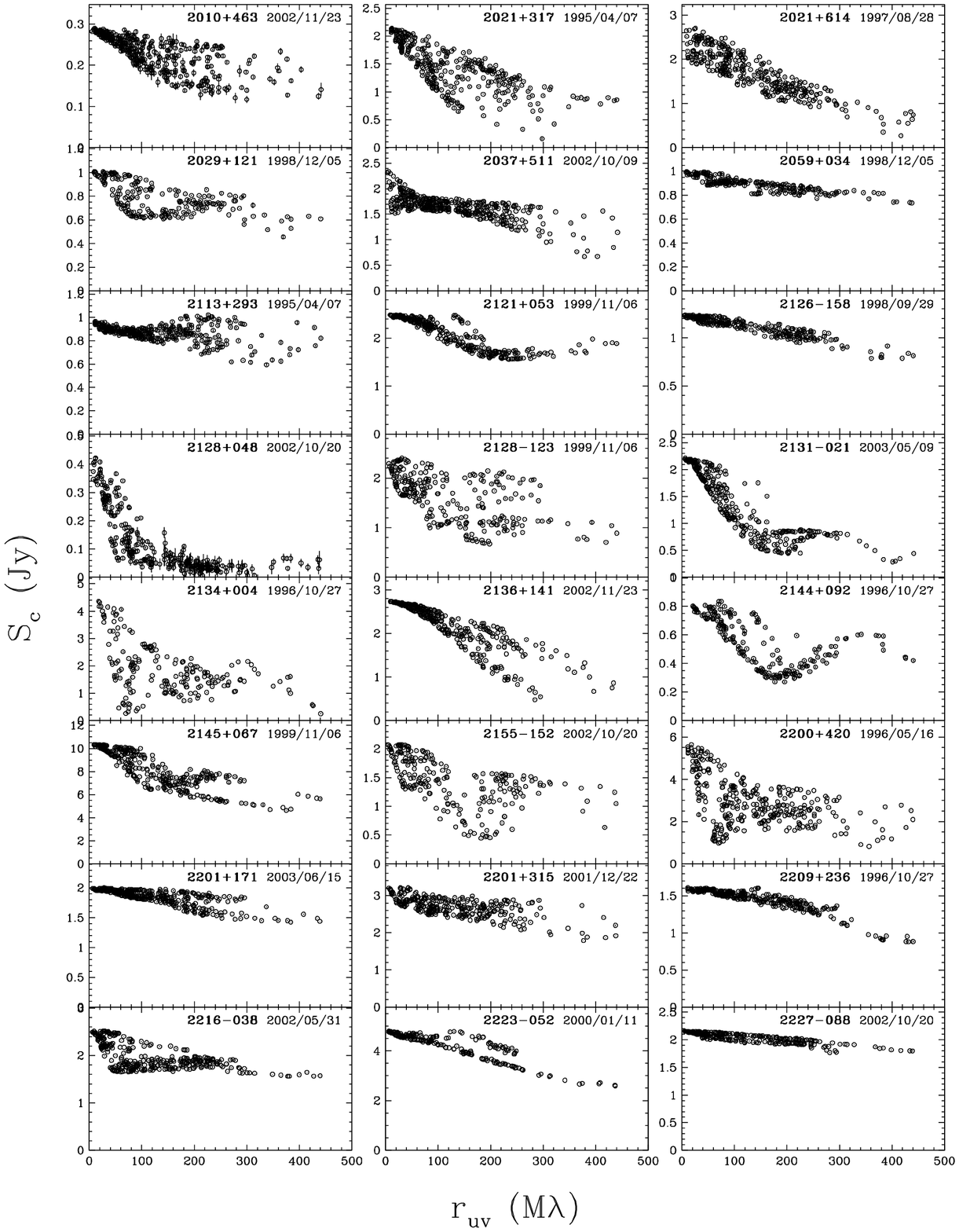}}
\caption{Continued}
\end{figure*}

\figurenum{\arabic{figure}}
\begin{figure*}[t]
\resizebox{1.04\hsize}{!}{\includegraphics[trim=0cm 12.5cm 0cm 1.5cm]{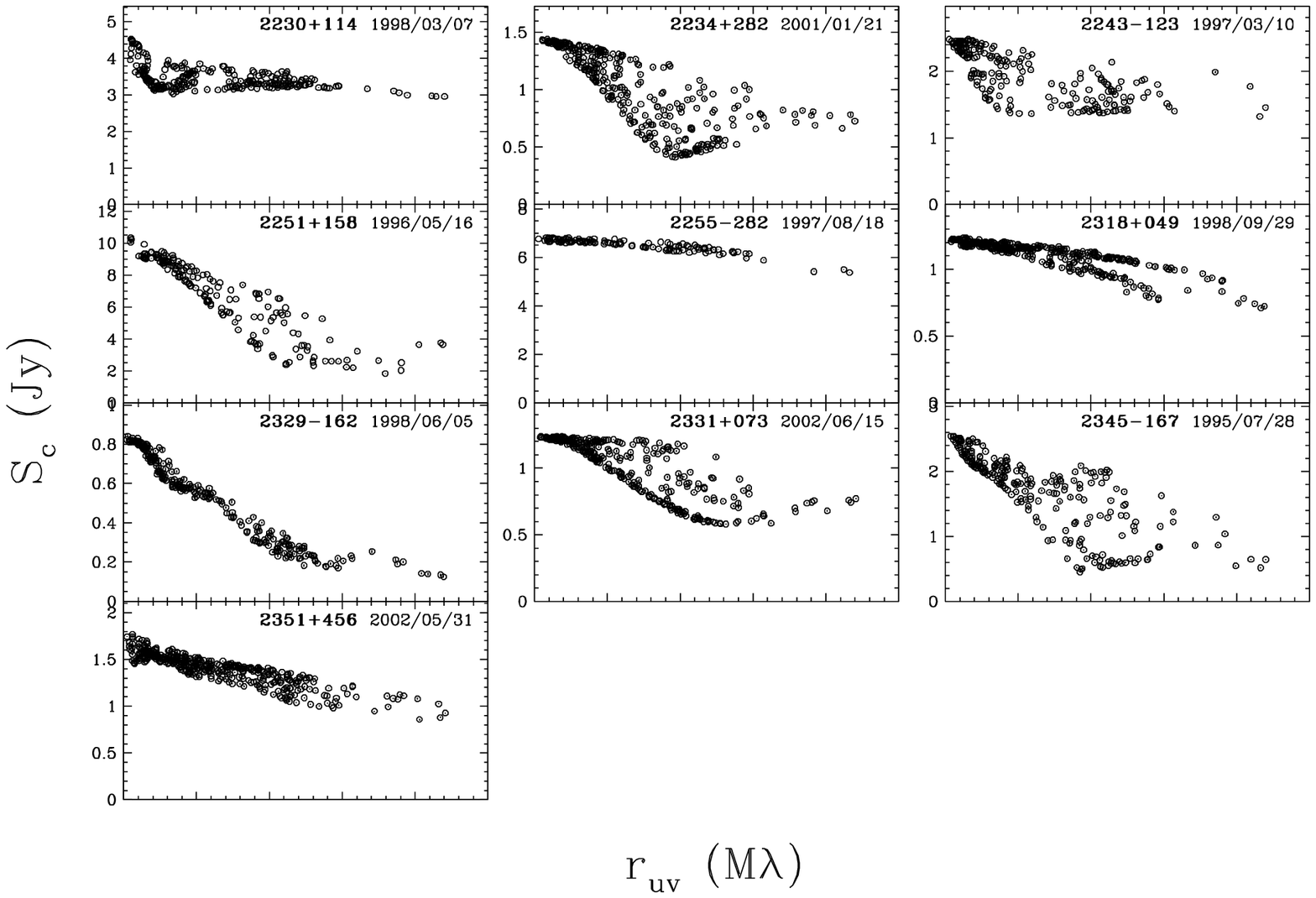}}
\caption{Continued}
\end{figure*}

\begin{figure*}[p]
\resizebox{1.04\hsize}{!}{\includegraphics[trim=0cm 1cm 0cm 1.5cm]{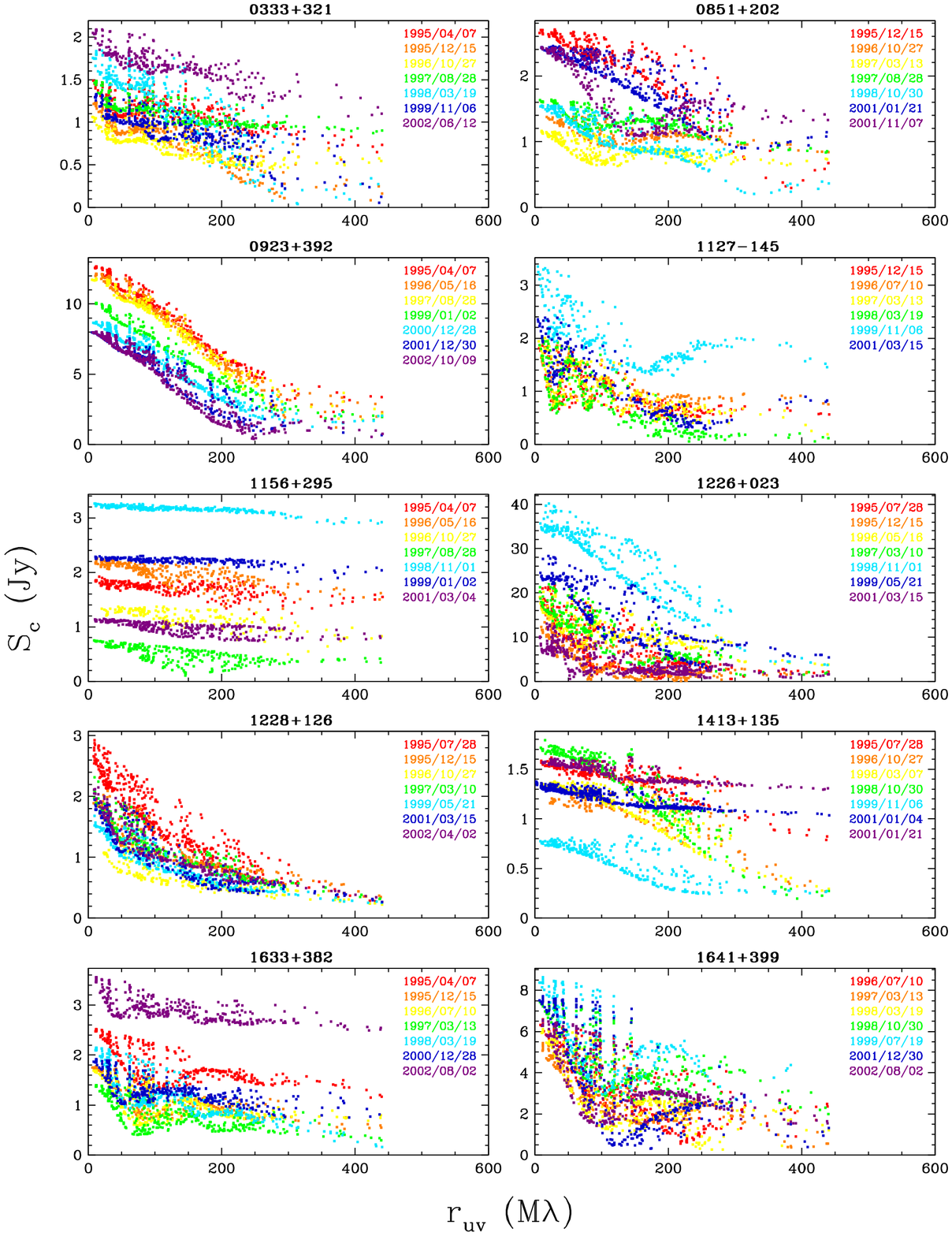}}
\caption{
Examples of the time variability of the correlated flux
density versus projected spacing for ten representative sources.
Different epochs are denoted by different colors in their
spectrum-color order.
}
\label{radplot_var}
\end{figure*}

     Our 15~GHz VLBA images, made with natural weighting of the
visibility data, have a nominal resolution of 0.5~mas in the east--west
direction and 0.6--1.3~mas in the north-south direction. The fringe
spacing of the VSOP Survey at 5~GHz
\citep{Lovell_etal2004,Scott_etal2004,Horiuchi_etal2004} is similar to
that of the VLBA at 15~GHz, but the effective resolution of the VLBA is
better, thanks to the relatively high SNR on the longest baselines, and
to the good relative calibration of the fringe visibilities, which can
be determined with self calibration using higher quality images based
on many more interferometer baselines, and full hour angle coverage.
Typically the dynamic range of the VLBA images (the ratio of the peak
flux density to the rms noise level) is better than 1000:1
(\citetalias{Kellermann_etal98,Zensus_etal02}, \citealt*{MOJAVE-I}).

     Figure~\ref{Auv} shows the visibility function amplitudes
(correlated flux density versus projected baseline length) for each
source in the full sample at the epoch when the amplitude is the
highest at the longest projected spacings \footnote{A version of
Figure~\ref{Auv} with plots for all of the epochs observed in the
15~GHz VLBA monitoring program until 2003/08/28 is published in the
electronic version only.}. These plots are independent of any
assumptions about the source structure, imaging artifacts, or beam
smoothing; and, more directly, they can show the presence of structure
on scales smaller than the synthesized beam. Also, as illustrated in
Figure~\ref{radplot_var}, many of these sources are variable. Changes
with time in the observed visibility data, especially those on the
longest baselines, corresponding to flux density variations in the
unresolved components, are not easily seen in the synthesized images
constructed from these data (see, e.g., Figure~\ref{radplot-map}), but
they are apparent when comparing visibility function plots. Variability
characteristics will be discussed in more detail in
\S~\ref{var_discussion}.

     Examination of the observed amplitudes of the visibility functions
in Figure~\ref{Auv} suggests that they can be divided into the
following categories:

     (i) Barely resolved sources where the fringe visibility decreases
only slowly with increasing spacing (e.g., 0235+164, 0716+714,
1726+455). For sources with good SNR, we can confidently determine that
these sources are resolved even if the fractional fringe visibility on
the longest baselines is as large as 0.95--0.98, which corresponds to
an angular size of only 0.056--0.036~mas in the direction corresponding
to the largest spacings (see the detailed discussion of the resolution
criterion in \S~\ref{modelfitting}). There are no sources which are
completely unresolved. However, the maximum resolution of the VLBA is
obtained within a narrow range of position angles close to the
east-west direction. In other directions, the resolution is poorer by a
factor of two to three.

     (ii) Sources with a well resolved component plus an unresolved or
barely resolved component. In these, the fringe visibility initially
decreases with increasing spacing, and then remains constant or
decreases slowly (e.g., 0106+013, 0923+392, 1213$-$172). For these
sources we can place comparable limits on the size of an unresolved
feature as in case (i) above.

     (iii) More complex or multi-component sources have visibility
functions which vary significantly with baseline. If there is an upper
envelope to the visibility function, which decreases only slowly to
larger spacings, then the structure is primarily one-dimensional, and
the upper envelope indicates the smallest dimension (e.g., 1045$-$188,
1538+149, 2007+777). If there is a well-defined lower envelope, which
monotonically decreases to larger spacings (e.g., 0014+813, 0917+624,
1656+053), this may be used as a measure of the overall dimensions of
the source. If minima are observed in the lower envelope (e.g.,
0224+671, 2131$-$021, 2234+282), they correspond to the spacing of the
major components.

\section{Derived Parameters and Model Fitting}
\label{modelfitting}

     The total flux density of each image, $S_\mathrm{VLBA}$, is the
sum of the flux densities of all components of the CLEAN model; this
should be equivalent to the visibility function amplitude,
$S_\mathrm{c}$ (the correlated flux density), on the shortest projected
baselines. In most cases in our sample, $S_\mathrm{c}$ at the shortest
spacings and $S_\mathrm{VLBA}$ are equal to within a few percent, which
is a consequence of the hybrid imaging procedure. We define the
($u$,$v$)-radius as $r_{uv}=\sqrt{u^2+v^2}$. The unresolved
(``compact'') flux density $S_\mathrm{unres}$ is defined as the upper
envelope (with 90\% of the visibilities below it) of the visibility
function amplitude $S_\mathrm{c}$ at projected baselines
$r_{uv}>$\,360\,M$\lambda$, which is approximately
0.8\,$r_{uv,\mathrm{max}}$. The overall uncertainty in
$S_\mathrm{VLBA}$ and $S_\mathrm{unres}$ is determined mainly by the
accuracy of the flux density (amplitude) calibration, which we estimate
to be about 5\% \citep[consistent with estimates of][]{Homan_etal02}.

     We have used the program DIFMAP~\citep{difmap} to fit the complex
visibility functions with simple models, consisting of two elliptical
Gaussian components, one representing the VLBA core, and the other the
inner part of a one-sided jet. By the ``core'' we mean the bright
unresolved feature typically found at the end of so-called ``core-jet''
sources; this is usually thought to be the base of a continuous jet,
and does not necessarily correspond to the nucleus of the object. The
main objective of our procedure was to obtain a robust characterization
of the core. We have verified the suitability of our method for that
purpose in several ways. Varying the initial values for the iterative
fitting procedure did not significantly change the final core parameter
values. Using more complex models consisting of three or four
components also did not significantly change the parameter values for
the core component in most sources, even fairly complex ones; instead,
the additional components tend to cover additional parts of the jets. We
have also compared the modeling results obtained in this study with the
more elaborate models obtained for all MOJAVE sources in the work of
\cite{MOJAVE-I}; those models were built up by adding new components
until the thermal noise level was reached in the residual image. For
about 90~\%\ of the sources in which the core was modeled by
\cite{MOJAVE-I} as an elliptical Gaussian component, the parameters
derived by the two methods agree to within 10~\%. However, for 23
sources with complex structure we found that a two-component model
overestimates the flux density and the angular size of the core, and we
have added more components to model these sources. We do not present or
use any modeling results for an additional 11 sources, which have very
complex structure, such as the two-sided radio galaxy
\objectname{NGC\,1052} (\citealt*{VRK03}, see also the sources in
Figure~2, \citetalias{Kellermann_etal04}). We conclude that, for most
of the sources in our sample, the core can be characterized accurately
and robustly with the two-component modeling method used, because the
beamed emission of the compact core dominates the 15~GHz structure
(median value of $S_\mathrm{core}/S_\mathrm{VLBA}=0.80$ for the full
sample).

     In about one quarter of the datasets the core is only
slightly resolved on the longest baselines. Following
\cite{Lobanov2005} we derive a resolution criterion for VLBI core
components, by considering a visibility distribution ${\cal
V}(r_\mathrm{uv})$ corresponding to the core. ${\cal V}(r_\mathrm{uv})$
is normalized by the flux density $S_\mathrm{core}$, so that ${\cal
V}(0)\equiv 1$. The core is resolved if
\begin{equation}
1 - {\cal V}(r_{uv,\mathrm{max}}) \le
\sigma_\mathrm{core}/S_\mathrm{core}=1/\mathrm{SNR}\,.
\label{eq:res-crit}
\end{equation}
Here, $\sigma_\mathrm{core}$ is the rms noise level in the area of
the image occupied by the core component (to exclude possible
contamination from the jet). In order to measure $\sigma_\mathrm{core}$
for each dataset, we have first subtracted the derived model from the
image and then used the residual pixel values in the area of the core
component convolved with the synthesized beam (truncated at the
half-power level). For naturally weighted VLBI data, the beam size,
with major and minor axes $b_\mathrm{maj}$ and $b_\mathrm{min}$
measured at the half-power point, yields the largest observed
($u$,$v$)-spacing as follows: $r_{uv,\mathrm{max}} =
(\pi\,b_\mathrm{maj}\,b_\mathrm{min})^{-1/2}$. The visibility
distribution corresponding to a Gaussian feature of angular size $d$ is
given by ${\cal V}(r_\mathrm{uv}) =
\exp[-(\pi\,d\,r_\mathrm{uv})2/(4\ln\,2)]$.  With these relations, the
resolution criterion given by equation (\ref{eq:res-crit}) yields for
the minimum resolvable size of a Gaussian component fitted to naturally
weighted VLBI data:
\begin{equation}
\theta_\mathrm{lim} = b_{\psi} \sqrt{ \frac{4 \ln 2}{\pi}
\ln\left(\frac{\mathrm{SNR}}{\mathrm{SNR}-1}\right) }\,.
\label{eq:resolution}
\end{equation}
Here, $b_{\psi}$ is the half power beam size measured along an
arbitrary position angle $\psi$. For all datasets we have derived
$\theta_\mathrm{lim}$ corresponding to the position angles of the major
and minor axis ($\theta_\mathrm{maj}$, $\theta_\mathrm{min}$) of the
fitted Gaussian core component. Whenever either one or both of the two
axes were smaller than the respective $\theta_\mathrm{lim}$, the
Gaussian component was considered to be
unresolved. $\theta_\mathrm{lim}$ was then used as an upper limit to
the size of the component, which yields a lower limit to its brightness
temperature. It should be noted that, at high SNR,
$\theta_\mathrm{lim}$ can be significantly smaller than the size of the
resolving beam and the Rayleigh limit. This is the result of applying a
specific {\em a priori} hypothesis about the shape of the emitting
region (a two-dimensional Gaussian, in our case) to fit the observed
brightness distribution. A similar approach is employed to provide the
theoretical basis for the technique of super resolution
\citep{BerteroDeMol96}.

     In our analysis we have also used the total flux density
$S_\mathrm{tot}$ at 15~GHz, determined from observations with single
antennas. We have incorporated the data from the University of Michigan
Radio Astronomy Observatory monitoring program
\citep[UMRAO,][]{Aller_etal85,AAH92,AAH03}\footnote{See also
{\sf{}\url{http://www.astro.lsa.umich.edu/obs/radiotel/umrao.html}}}
as well as instantaneous 1--22~GHz broad-band radio spectra obtained
during the long-term monitoring of compact extragalactic sources with
the \mbox{RATAN--600} radio telescope of the Special Astrophysical
Observatory \citep{Kovalev98,Kovalev_etal99,Kovalev_etal2000}. We have
interpolated the UMRAO observations in time, and the \mbox{RATAN--600}
data in both frequency (between 11 and 22~GHz) and time to obtain the
effective filled aperture total flux density, $S_\mathrm{tot}$, at the
epoch and frequency of the VLBA observations. The main contribution to
the total uncertainty on the $S_\mathrm{tot}$ values comes from the
non-simultaneity of the VLBA and the single dish observations. This can
give errors up to 20--30\%, but the typical uncertainties are below 5\%.

\section{Results and Discussion}
\label{results}
\subsection{Source Compactness}

\begin{figure}[t]
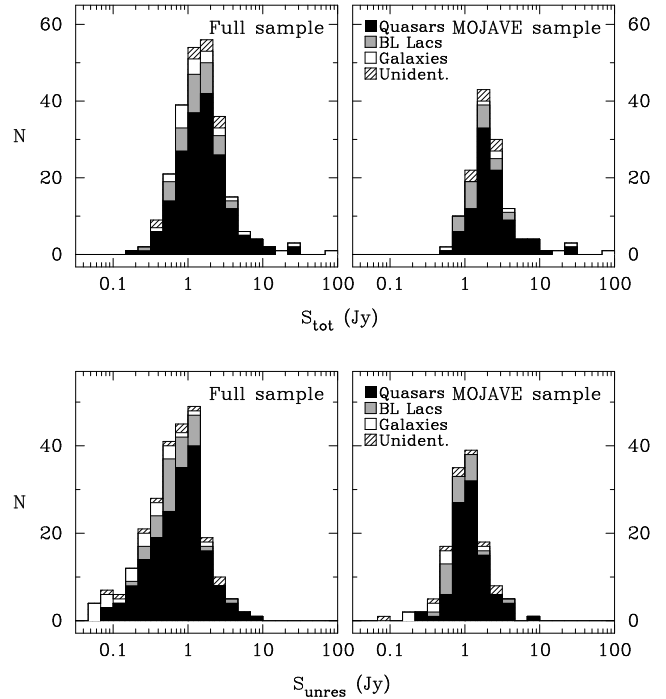

\begin{center}
\resizebox{1.0\hsize}{!}{
\includegraphics[trim=-0.2cm 0cm 0cm 0cm]{f4a.eps}
}
\resizebox{1.0\hsize}{!}{
\includegraphics[trim=-0.2cm 0cm 0cm -1cm]{f4b.eps}
}
\end{center}
\caption{Distribution of the flux density in the full and MOJAVE samples
showing the median single antenna flux density $S_\mathrm{tot}$ (upper panel),
and the median flux density of the most compact component
$S_\mathrm{unres}$ (lower panel).
}
\label{hist_s}
\end{figure}

\begin{figure}[t]
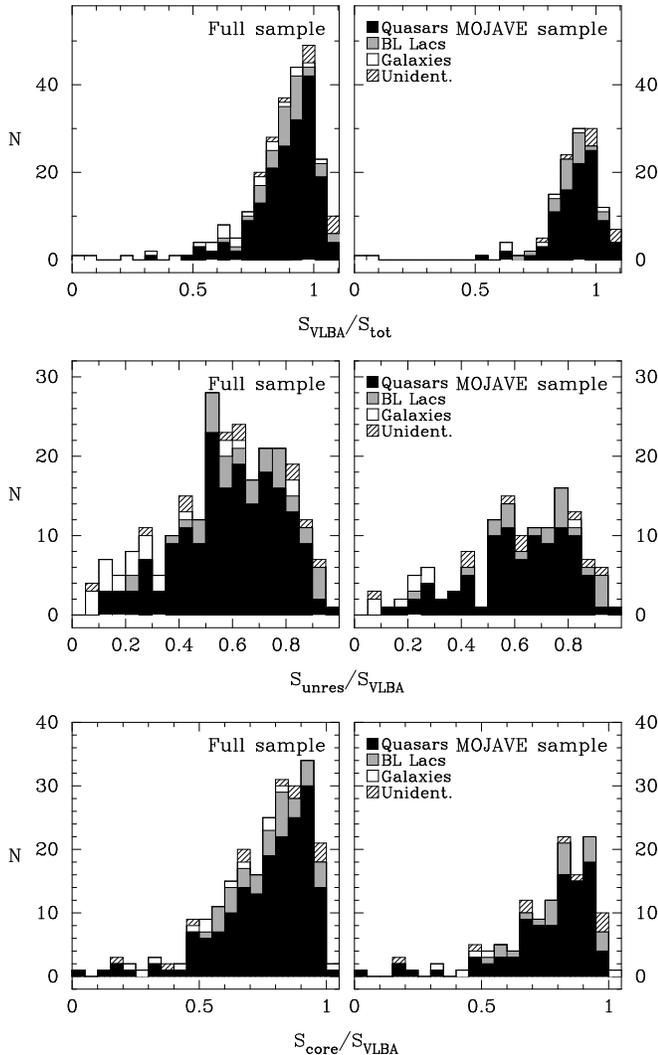

\begin{center}
\resizebox{1.0\hsize}{!}{\includegraphics[trim=0cm   0cm 0cm 0cm]{f5a.eps}}
\resizebox{1.0\hsize}{!}{\includegraphics[trim=0cm   0cm 0cm -0.5cm]{f5b.eps}}
\resizebox{1.0\hsize}{!}{\includegraphics[trim=0cm 0.7cm 0cm -0.5cm]{f5c.eps}}
\end{center}
\caption{
Distributions of the median compactness indices on arcsecond scales
$S_\mathrm{VLBA}/S_\mathrm{tot}$ (upper panel) and on
sub-milliarcsecond scales $S_\mathrm{unres}/S_\mathrm{VLBA}$ (middle panel)
as well as the VLBA core dominance
$S_\mathrm{core}/S_\mathrm{VLBA}$ (lower panel).
A few sources have an apparent compactness index at arcsecond scales
$S_\mathrm{VLBA}/S_\mathrm{tot}>1$; this is due to source variability
and the non-simultaneity of the VLBA and single antenna observations.
}
\label{hist_ss}
\end{figure}

     Figure~\ref{hist_s} shows the distributions in our sample of the
total flux densities, $S_\mathrm{tot}$, from single dish measurements,
and the correlated flux densities, $S_\mathrm{unres}$, from long VLBA
spacings. The peak in the distribution of $S_\mathrm{tot}$ corresponds
to our nominal flux density limit of 1.5 or 2~Jy (depending on
declination). The tail to lower flux densities in both panels is due to
variability, and in the full sample (left hand panel) the tail also
includes some sources of particular interest, which we included in the
observations, but which did not meet our flux density criteria.
     Figure~\ref{hist_ss} gives the distributions of the ``indices of
compactness'' on arcsecond scales, $S_\mathrm{VLBA}/S_\mathrm{tot}$, and
sub-mas scales, $S_\mathrm{unres}/S_\mathrm{VLBA}$, as well as of the
VLBA core dominance, $S_\mathrm{core}/S_\mathrm{VLBA}$.

     Many sources in our sample have considerable flux on spatial
scales sampled by the longest VLBA baselines. In the lower left hand
panel of Figure~\ref{hist_s}, we see that more than 90\% of the sources
have an unresolved flux density greater than 0.1~Jy at projected
baselines longer than 360 million wavelengths, while the middle left
panel of Figure~\ref{hist_ss} shows that 68\% of the sources have a
median $S_\mathrm{unres}/S_\mathrm{VLBA}>0.5$. Table~\ref{data_table}
lists, for each source, flux densities and model fitting results (as
well as some other data) at the epoch for which its unresolved flux
density $S_\mathrm{unres}$ was greatest. These data will be of value
for various purposes, including planning future VLBI observations using
Earth-space baselines. For 163 of these sources, the median flux
density of the most compact component is greater than 0.5~Jy.

     We have compared the measured values of $S_\mathrm{VLBA}$ and
$S_\mathrm{tot}$. Figure~\ref{hist_ss} indicates that there are no
significant systematic errors in the independently-constructed
VLBA/RATAN/UMRAO flux density scales. The median compactness index on
arcsecond scales, $S_\mathrm{VLBA}/S_\mathrm{tot}$, is 0.91 for the
full sample and 0.93 for the MOJAVE sample, which indicates that for
most sources the VLBA image contains nearly all of the flux density.
Some sources have an apparent compactness on arcsecond scales
$S_\mathrm{VLBA}/S_\mathrm{tot}>1$. Most likely, this is due to source
variability and the non-simultaneity of the VLBA and single antenna
observations. Sources with compactness index close to unity (see
Table~\ref{data_table}) are well-suited as calibrators for other VLBA
observations.
     
\subsection{\label{source_classes}Source Classes}

\begin{figure}[t]
\resizebox{1.0\hsize}{!}{
\includegraphics[trim=-0.5cm 0cm 0cm -0.5cm]{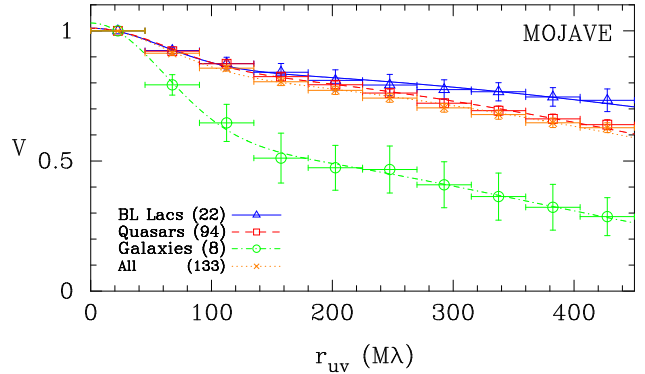}
}
\caption{
The non-weighted mean of the fringe visibility
versus projected spacing for the MOJAVE sample.
The distribution is normalized to 1.0 at 25 million wavelengths.
Visibility data at the epoch with the maximum correlated
flux density at maximum VLBA spacings for each source is used.
Before averaging over the samples, the fringe visibility amplitude
for each source was binned and averaged.
Vertical bars represent the $\pm1\sigma$ formal error,
horizontal bars~--- intervals of visibility data (ten bins
in total).
The lines represent the best fitting two-component models.
See model parameters in Table~\ref{radplot-model}.
}
\label{relative_radplot}
\end{figure}

     The curves in Figure~\ref{relative_radplot} show the mean
visibility amplitude versus projected ($u$,$v$) spacing, averaged over
all sources in the MOJAVE sample, and averaged over the MOJAVE quasars,
BL~Lacs, and active galaxies, separately. The best fitting parameter
values for a model consisting of two Gaussian components are listed in
Table~\ref{radplot-model} for each of these mean visibility curves.

     The active galaxies are, on average, the least VLBA core dominated and
the least compact on arcsecond (Figure~\ref{hist_ss}) and sub-mas
(Figures~\ref{hist_ss} and~\ref{relative_radplot}, and
Table~\ref{radplot-model}) scales. The fact that the relative
contribution of an extended component (i.e.\ a jet) is significantly
greater for active galaxies is consistent with unification models in
which radio galaxies are viewed at larger angles to the line of sight
than BL~Lacs or quasars
\citep[e.g.,][]{Antonucci_etal87,Antonucci93,UrryPadovani95,Wills99},
so that the latter have higher Doppler factors, their cores are more
boosted, and thus they appear more core dominated. The
Kolmogorov-Smirnov (K-S) test confirms that for both the full and the
MOJAVE sample the probability is less than 1\%\ that the active
galaxies have the same parent distribution as the quasars or the
BL~Lacs with regard to their compactness on arcsecond scales,
$S_\mathrm{VLBA}/S_\mathrm{tot}$, or their compactness on sub-mas
scales, $S_\mathrm{unres}/S_\mathrm{VLBA}$; with regard to their core
dominance, $S_\mathrm{core}/S_\mathrm{VLBA}$, the probability is less
than 2\%. 

     For the BL~Lacs versus quasars, K-S tests were inconclusive.
However, Figure~\ref{relative_radplot} and Table~\ref{radplot-model}
show that the BL~Lacs are, on average, even more compact on sub-mas
scales than the quasars. We have also found this distinction between
sub-samples of quasars and BL~Lacs chosen to have statistically
indistinguishable redshift distributions. The differences between the
quasars and the BL~Lacs in angular size at sub-mas scales in the sample
as a whole are therefore not related to the different overall redshift
distributions of these groups. As discussed in \S~\ref{sampledef},
classifying objects is a complex issue, particularly with regard to
BL~Lacs. With our tabulated data, others could repeat the analysis
using their own classification procedure if desired. However, the
optical classification scheme we have used is evidently ``clean''
enough that, after the fact, it turns out to correspond to differences
in radio compactness. We cannot image any hypothetical optical
classification bias which could be fully responsible for the
correspondence with radio compactness, and we conclude that it is an
actual physical phenomenon.


     Our sample does not show a significant dependence on redshift of
the index of compactness on sub-mas scales,
$S_\mathrm{unres}/S_\mathrm{VLBA}$, although the few heavily resolved
sources are mostly active galaxies at low redshift.

\subsection{Frequency Dependence}  

     \cite{Horiuchi_etal2004} have presented a plot, similar to
Figure~\ref{relative_radplot}, based on 5~GHz VLBA and VSOP
observations of 189 radio sources which cover a comparable range of
spatial frequencies as our 15~GHz VLBA data. They find that the average
fringe visibility in the range 400 to 440 million
wavelengths is 0.21--0.24. For the 116 sources in common to the two
samples (see Table~\ref{sample}), we find an average fringe visibility
at 15~GHz of about 0.6. The compact component emission dominates at
15~GHz ($>$75\%, see Table~\ref{radplot-model}), but not at 5~GHz
\citep[40\%,][]{Horiuchi_etal2004}. This reflects the fact that the
5~GHz observations detect a larger contribution from steep spectrum
optically thin large scale components. The 5~GHz VSOP survey sample and
our 15~GHz VLBA sample are not identical, but this result is confirmed
if only the subset of overlapping Pearson--Readhead VSOP survey sources
\citep{Lister_etal01,Horiuchi_etal2004} is used for comparison.

\subsection{Brightness Temperatures}

\begin{figure}[t]
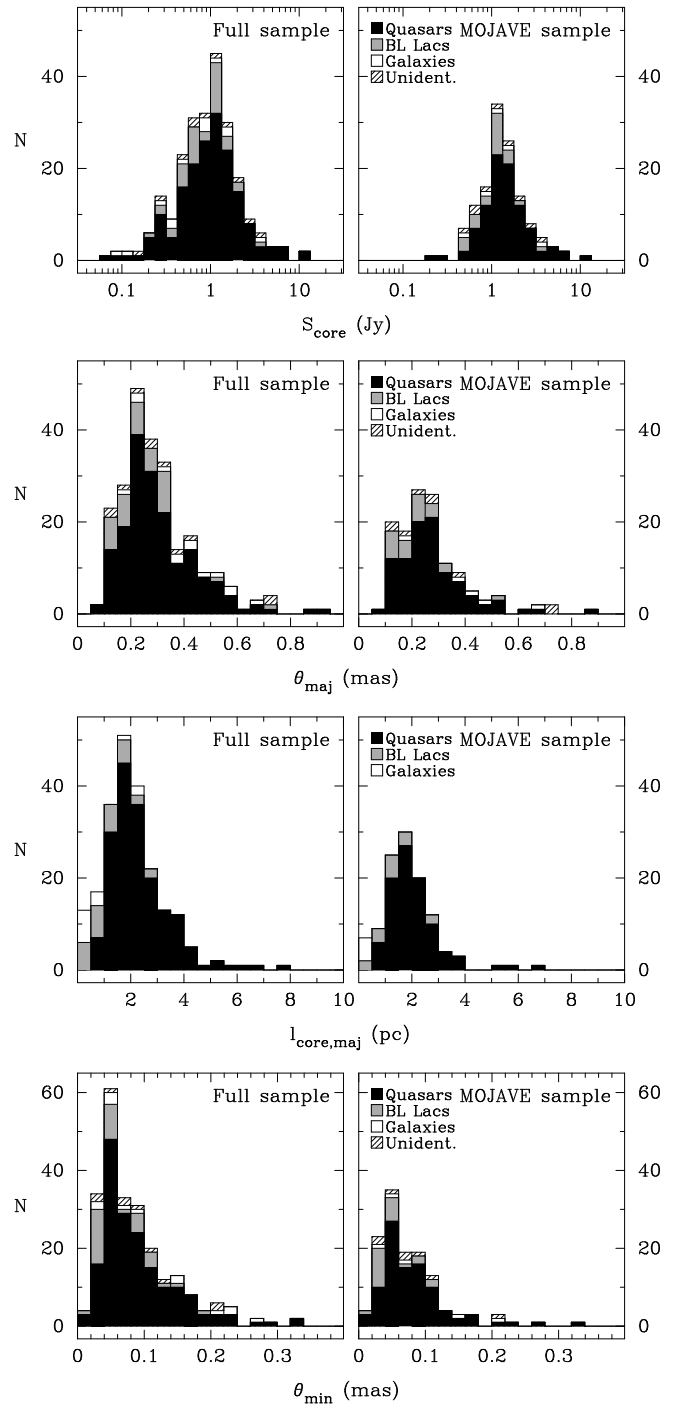

\begin{center}
\resizebox{1.0\hsize}{!}{
\includegraphics[trim=0cm -0.5cm 0cm 0cm]{f7a.eps}}
\resizebox{1.0\hsize}{!}{
\includegraphics[trim=0cm -0.5cm 0cm 0cm]{f7b.eps}}
\resizebox{1.0\hsize}{!}{
\includegraphics[trim=0cm -0.5cm 0cm 0cm]{f7c.eps}}
\resizebox{1.0\hsize}{!}{
\includegraphics[trim=0cm +0.7cm 0cm 0cm]{f7d.eps}}
\end{center}
\caption{From top to bottom,
the distributions of the median flux density, angular
and linear dimensions of core model components derived from the
multi-epoch observations for each source.
}
\label{ft_difmap}
\end{figure}

\begin{figure}[t]
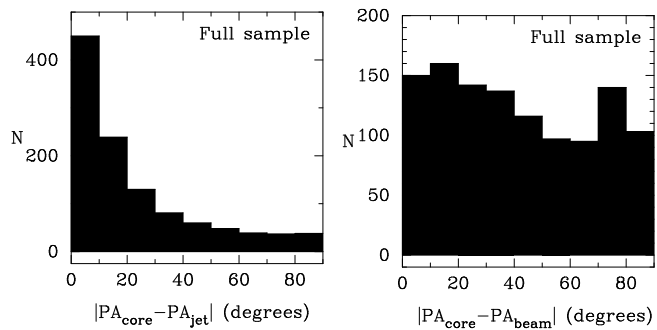

\begin{center}
\resizebox{1.0\hsize}{!}{
\includegraphics[trim=   0cm 0cm 0cm 0cm]{f8a.eps}
\includegraphics[trim=-0.5cm 0cm 0cm 0cm]{f8b.eps}
}
\end{center}
\caption{Distributions of the difference between the
position angle of the major axis of the core and the
position angle of a jet (left) and the difference between the
position angle of the major axis of the core and the position
angle of the major axis of the VLBA beam (right) for all the
data modeled.
}
\label{hist_PABJCdiff}
\end{figure}

     Figure~\ref{ft_difmap} shows distributions of the core parameters.
The ratio of the major axis of the core to the beam width in the same
direction varies by more than an order of magnitude, so, in most cases,
we believe that our measured core dimensions are not an artifact of
the finite beam size.

     The cores are always resolved along their major axis. However, for
158 sources in our full sample, there is at least one epoch at which the
core component appears unresolved along the minor axis, where it is
then typically less than 0.05~mas in size. In 19 of these sources,
including 5 BL~Lacs, the core is unresolved along the minor axis at all
observed epochs.

     Figure~\ref{hist_PABJCdiff} shows the distribution of the
difference between the position angle of the major axis of the core,
$\mathrm{PA}_\mathrm{core}$, and the jet direction,
$\mathrm{PA}_\mathrm{jet}$; the latter was taken to be the median over
all epochs of the position angle of the jet component with respect
to the core. All possible values of
$|\mathrm{PA_{core}}-\mathrm{PA_{jet}}|$ between $0^\circ$ and
$90^\circ$ are observed. Not unexpectedly, the peak of the distribution
is close to zero; that is, the Gaussian component representing the core
is typically extended along the jet direction. For the majority of
sources which have multi-epoch modeling data, the orientation of the
core is stable in time, with a scatter around the average
$\mathrm{PA}_\mathrm{core}$ of less than $10^\circ$.
Figure~\ref{hist_PABJCdiff} also demonstrates that the position angle of
the core is not correlated with the position angle of the VLBA beam, so
that the measured core orientation is, in most cases, not distorted by
the orientation of the VLBA beam.

The brightness temperature of a slightly resolved component in the
rest frame of the source is given by
\begin{equation}
T_\mathrm{b}= 
\frac{2\,\ln2}{\pi k}\frac{S_\mathrm{core}\,\lambda^2\,(1+z)}
{\theta_\mathrm{maj}\,\theta_\mathrm{min}}\,,
\label{Tb_eqn}
\end{equation}
where $S_\mathrm{core}$ is the flux density of a VLBA core,
$\theta_\mathrm{maj}$ and $\theta_\mathrm{min}$ are the
full width at half maximum (FWHM)
of an elliptical Gaussian component along the major and
the minor axis, $\lambda$ is the wavelength of
observation, $z$ is the redshift, and $k$ is the Boltzmann constant.
Observing at $\lambda=2$~cm with $S_\mathrm{core}$ measured in Jy,
and $\theta_\mathrm{maj}$ and $\theta_\mathrm{min}$ in mas, we can write
\begin{equation}
T_\mathrm{b}= 
5.44\,\times\,10^9\,\frac{S_\mathrm{core}\,(1+z)}{\theta_\mathrm{maj}\,\theta_\mathrm{min}}
\quad\mathrm{K}\,.
\end{equation}
The brightness temperature can also be represented in terms of an
effective baseline
$D=\lambda/\sqrt{\theta_\mathrm{maj}\theta_\mathrm{min}}$. If $D$ is
measured in km and $S_\mathrm{core}$ in Jy, we have
\begin{equation}
T_\mathrm{b}= 
3.20\,\times\,10^{2}\,S_\mathrm{core}\,D^2\,(1+z)\quad\mathrm{K}\,,
\label{Tb_eqn_notheta}
\end{equation}
which is independent of wavelength and depends only on the physical
length of the effective projected baseline and on the core flux density.
For sources without measured redshift (see Table~\ref{sample})
we use $z=0$ to define a limit to $T_\mathrm{b}$.

\begin{figure}[t]
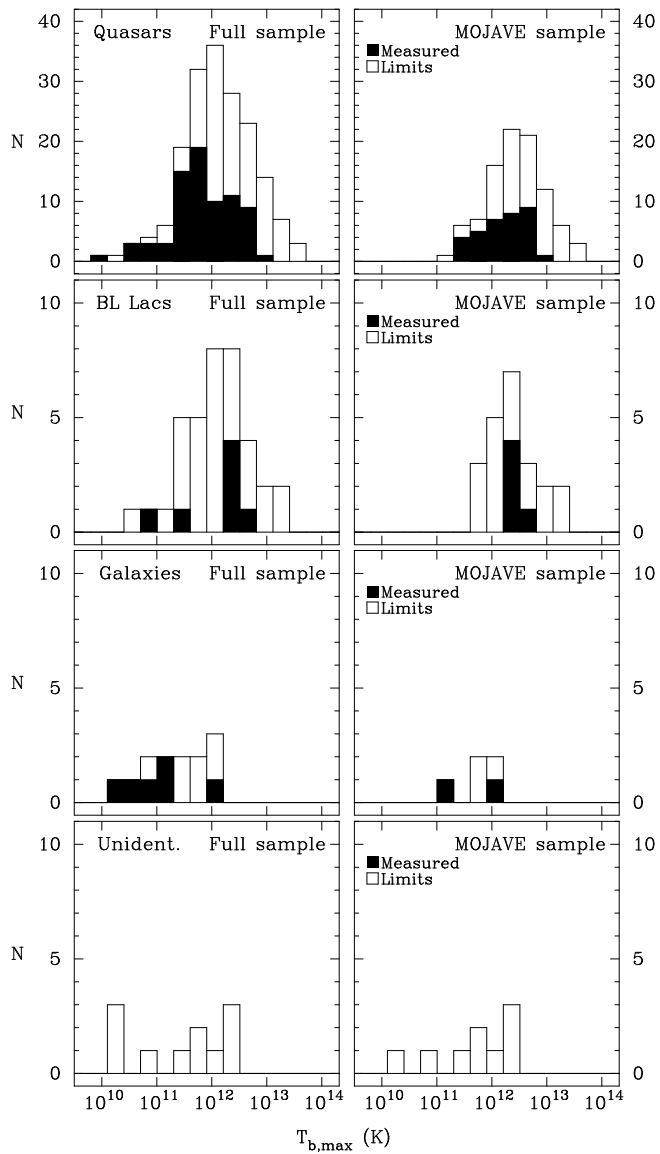

\begin{center}
\resizebox{1.0\hsize}{!}{
\includegraphics[trim=0cm 1.80cm 0cm 0cm,clip]{f9a.eps}
}
\resizebox{1.0\hsize}{!}{
\includegraphics[trim=0cm 1.80cm 0cm 0cm,clip]{f9b.eps}
}
\resizebox{1.0\hsize}{!}{
\includegraphics[trim=0cm 1.80cm 0cm 0cm,clip]{f9c.eps}
}
\resizebox{1.0\hsize}{!}{
\includegraphics[trim=0cm 0cm 0cm 0cm]{f9d.eps}
}
\end{center}
\caption{
Distributions of the maximum observed VLBA core brightness
temperature in the source frame. Each bin covers a factor
of two in brightness temperature.
}
\label{hist_Tb}
\end{figure}

     \citetalias{Kellermann_etal98} and \citetalias{Zensus_etal02} gave
conservative estimates of the observed peak brightness temperature
based on the observed angular size, which is the intrinsic size
convolved with the VLBA beam width. Here, we derive the core brightness
temperature using the dimensions or upper limits obtained from direct
modeling of the complex visibility functions. For many sources, the
effective resolution is an order of magnitude better than given by
\citetalias{Kellermann_etal98} and \citetalias{Zensus_etal02}, so the
corresponding derived brightness temperatures are as much as a factor
of 100 greater. The median value of these VLBA core brightness
temperatures, shown in Figure~\ref{hist_Tb}, is near $10^{12}$~K; they
extend up to $5\,\times\,10^{13}$~K. This is comparable with brightness
temperatures derived from VSOP space VLBI observations
\citep{Hirabayashi_etal00,Frey_etal00,Tingay_etal01,Horiuchi_etal2004}.
In many cases our measurement refers only to the upper limit of the
angular size, corresponding to our minimum resolvable size derived
using equation (\ref{eq:resolution}). The effective resolution depends
on the maximum baseline and on the signal-to-noise ratio near the maximum
resolution. The true brightness temperatures of many sources may extend
to a much higher value, beyond the equipartition value of $10^{11}$~K
\citep{Readhead94,SG85} or the inverse Compton limit of
$10^{12}$~K~\citep{Kellermann_Pauliny-Toth69}. These high brightness
temperatures are probably due to Doppler boosting, but transient
non-equilibrium events, coherent emission, emission by relativistic
protons, or a combination of these effects
\citep[e.g.,][]{Kardashev00,Kellermann02,Kellermann03} may also play a
role.

\begin{figure}
\begin{center}
\resizebox{1.0\hsize}{!}{
\includegraphics[trim=0cm 0cm 0cm 0cm]{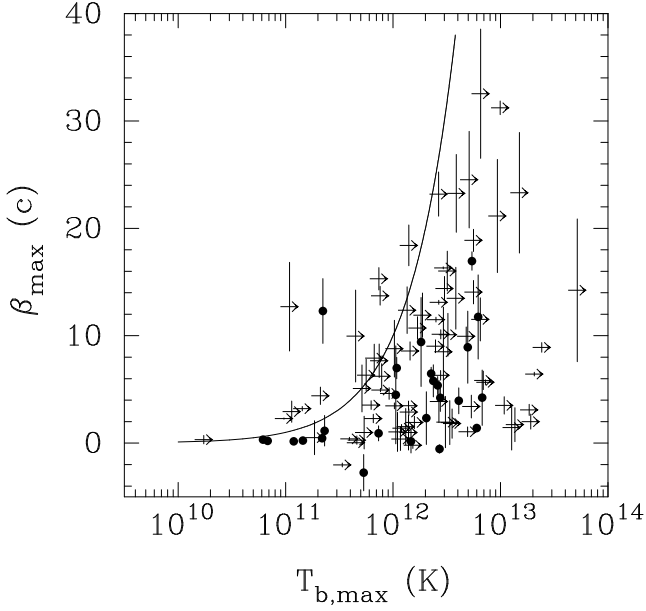}
}
\end{center}
\caption{Apparent velocity, $\beta_\mathrm{max}$, from \citetalias{Kellermann_etal04}
versus maximum core brightness temperature, $T_\mathrm{b,max}$.
Values of $\beta_\mathrm{max}$ are taken for the fastest components with
ratings `E' or `G' only.
Lower limits of brightness temperature are plotted as arrows.
The curve is plotted for $T_\mathrm{b,max}=\beta_\mathrm{max} T_\mathrm{int}$,
where the intrinsic brightness temperature is taken as $T_\mathrm{int}=10^{11}$\,K.
}
\label{beta_maxTb}
\end{figure}

     If the high observed brightness temperatures are due to Doppler
boosting, and if the range of intrinsic brightness temperatures,
$T_\mathrm{int}$, is small, there should be a correlation between the
apparent jet velocity, $\beta_\mathrm{app}$, and the observed
brightness temperature. For those sources listed in
\citetalias{Kellermann_etal04}, Figure~\ref{beta_maxTb} shows the
fastest observed jet velocity against the maximum observed brightness
temperature of their core. While this plot contains mostly lower limits
to the brightness temperature, there are no sources with a low
brightness temperature and a high observed speed; conversely, the
highest speeds are observed only in sources with a high brightness
temperature. This is the trend which we would expect if the observed
brightness temperatures are Doppler boosted with $T_\mathrm{obs} =
\delta T_\mathrm{int}$, where $\delta$ is the Doppler factor. At the
optimum angle $\vartheta$ to the observer's line of sight for superluminal
motion, $\beta =\cos\vartheta$, $\beta_\mathrm{app} = \delta$ and
therefore $T_\mathrm{obs} \simeq \beta_\mathrm{app}T_\mathrm{int}$. As
shown in Figure~\ref{beta_maxTb}, with $T_\mathrm{int} = 10^{11}$\,K,
this curve tracks the trend of the data. Of course, the actual jet
orientations deviate from the optimum viewing angle given by
$\beta=\cos\vartheta$, and many of our brightness temperature estimates
are lower limits. Both of these factors lead to a spread in the data,
so we should not expect a tight correlation along the plotted line;
however, the general agreement between the trend of the data and this
simple model supports the idea that the intrinsic brightness
 temperatures have been Doppler boosted by the same relativistic motion
that gives us the observed component speeds. We note that there are
some sources with high brightness temperatures but low speeds. This is
expected, as some sources will have an angle to the line of sight much
smaller than $\beta=\cos\vartheta$, and those sources will have a small
apparent motion but will still be highly beamed (M.~H.~Cohen et al., in
preparation).

     For those sources where there are multiple epochs of observation,
the core parameters, in particular the observed brightness
temperatures, vary significantly with time. Population modeling of the
distribution of brightness temperature, as well as comparisons with the
results of other VLBI surveys \citep[e.g.,][]{Lobanov_etal00} may give
insight into the distributions of intrinsic brightness temperatures and
Doppler factors.

     We have not found any significant correlation between redshift and
brightness temperature. This is in agreement with the 5~GHz VSOP
results of \cite{Horiuchi_etal2004}.

\subsection{Variability}
\label{var_discussion}

     The long term cm-wave monitoring data on our sources from UMRAO and
RATAN show complex light curves with frequent flux density outbursts
\citep[e.g.,][]{AAH03,KKNB02}. These outbursts are thought to be
associated with the birth of new compact features, which are often not
apparent in VLBI images until they have moved sufficiently far down the
jet. Changes usually appear sooner in the visibility function
(Figure~\ref{radplot_var}), which in the case of our data, probes
angular scales roughly 10 times smaller than the typical image
restoring beam.

Most new jet features typically increase in size and/or decrease in
flux density after a few months to years as a result of adiabatic
expansion and/or synchrotron losses. However, an interesting exception
is M\,87 (1228+126), where the most compact feature appears to remain
constant (Figure~\ref{radplot_var}), although the larger scale jet
structure shows changes by up to a factor of two in correlated flux
density. This unusual behavior, which was first noted by
\cite{Kellermann_etal73}, is remarkable in that the dimensions of this
compact stable feature are only of the order of ten lightdays or
less. This weak ($\lesssim 0.2$~Jy) stable feature in the center of
\objectname{M\,87} may be closely associated with the accretion
region. More sensitive observations with comparable linear resolution
might show similar phenomena in more distant sources, but such
observations will only be possible with large antennas in space.

\begin{figure}
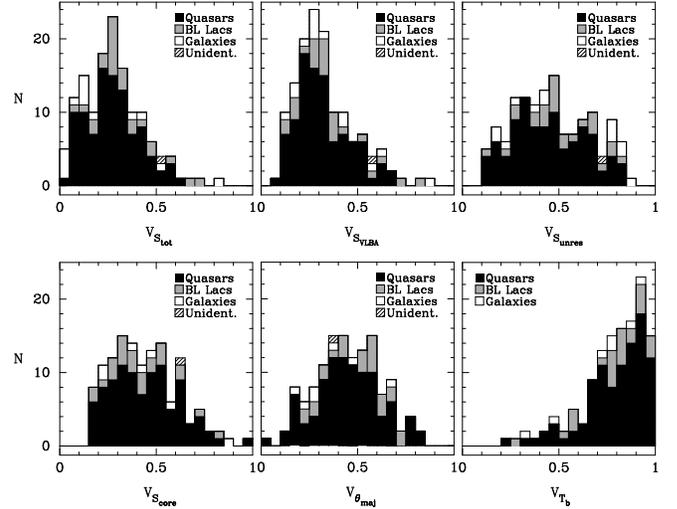

\resizebox{1.0\hsize}{!}{
\includegraphics[trim=0cm   0cm 0cm 0cm]{f11a.eps}
\includegraphics[trim=1.7cm 0cm 0cm 0cm,clip]{f11b.eps}
\includegraphics[trim=1.7cm 0cm 0cm 0cm,clip]{f11c.eps}
}
\resizebox{1.0\hsize}{!}{
\includegraphics[trim=0cm       0cm 0cm -0.5cm]{f11d.eps}
\includegraphics[trim=1.7cm  0.06cm 0cm -0.5cm,clip]{f11e.eps}
\includegraphics[trim=1.7cm     0cm 0cm -0.5cm,clip]{f11f.eps}
}
\caption{
Distributions of the variability index, $V$, derived for
$S_\mathrm{tot}$, $S_\mathrm{VLBA}$, $S_\mathrm{unres}$,
core flux density, major axis, and brightness temperature.
The variability indices are calculated
for 137 sources of the full sample observed four or more times.
}
\label{Vindex}
\end{figure}

     We define a variability index as
$V_X=\frac{X_\mathrm{max}-X_\mathrm{min}}{X_\mathrm{max}+X_\mathrm{min}}$.
Figure~\ref{Vindex} shows distributions of the variability indices
$V_{S_\mathrm{tot}}$, $V_{S_\mathrm{VLBA}}$, and $V_{S_\mathrm{unres}}$,
as well as, to represent the cores, 
$V_{S_\mathrm{core}}$, $V_{\theta_\mathrm{maj}}$, and $V_{T_\mathrm{b}}$.
As expected, the variability indices for the flux density become larger
with improved resolution, going from median values of
$V_{S_\mathrm{tot}}=0.27$ and $V_{S_\mathrm{VLBA}}=0.30$ to
$V_{S_\mathrm{core}}=0.42$ and $V_{S_\mathrm{unres}}=0.45$.  For about
68\% of the sources, the flux density of the core has varied by a
factor of 2 or more, that is $V > 1/3$. Similarly, the size of the core
major axis, $\theta_\mathrm{maj}$, changed by as much as a factor of 5,
or $V > 2/3$, in some cases. Probably, this is due to the creation or
ejection of a new component, which initially is not resolved from
the core, but then separates from it, causing first an
apparent increase, and then an apparent decrease in the strength and
size of the core. The observed strong variability of the brightness
temperatures (median $V_{T_\mathrm{b}}=0.82$) may reflect strong
variations of particle density (due to ejections) and/or magnetic field
strength. The most variable sources tend to have the most compact
structure. A variability index $V_{S_\mathrm{VLBA}}>0.6$ is observed
for nine sources, all but one of which have sub-mas compactness
$S_\mathrm{unres}/S_\mathrm{VLBA}\ge0.75$.

\subsection{Intra-Day Variable Sources}

\begin{figure}
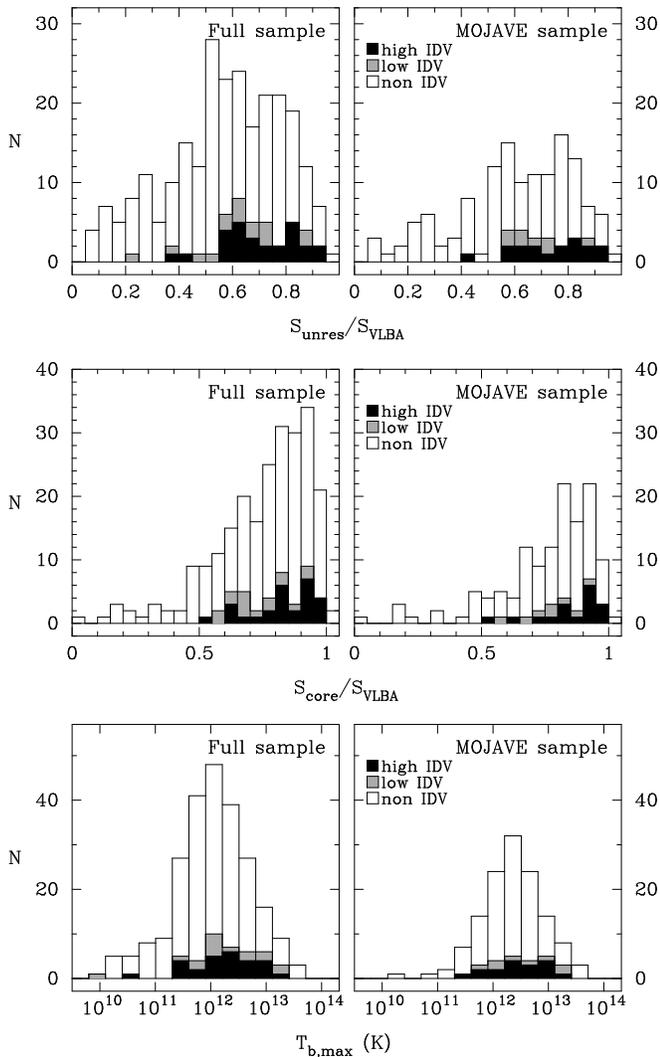

\begin{center}
\resizebox{1.0\hsize}{!}{\includegraphics[trim=0cm 0cm 0cm 0cm]{f12a.eps}}
\resizebox{1.0\hsize}{!}{\includegraphics[trim=0cm 0cm 0cm -0.5cm]{f12b.eps}}
\resizebox{1.0\hsize}{!}{\includegraphics[trim=0cm 1cm 0cm -0.5cm]{f12c.eps}}
\end{center}
\caption{
Distributions of the median compactness index on
sub-milliarcsecond scales $S_\mathrm{unres}/S_\mathrm{VLBA}$ (upper
panels), the VLBA core dominance $S_\mathrm{core}/S_\mathrm{VLBA}$
(middle panels) and maximum brightness temperature $T_\mathrm{b,max}$
(lower panels) for IDV selected and non-selected sources. We separate
IDVs with high modulation index ($m>0.02$ observed at least once ---
``high IDV'') from ones with low modulation index (measured values of
modulation index always less than 0.02 or are not reported --- ``low
IDV'').
}
\label{IDV}
\end{figure}

     We have used the results of several IDV search and monitoring
programs at the Effelsberg 100~meter telescope, the VLA, and the ATCA
at 1.4~to 15~GHz \citep{Quirrenbach_etal92, Quirrenbach_etal2000,
KC_etal01, Bignall_etal02, Kraus_etal03, Lovell_etal03}, to identify
IDV sources in our sample (Table~\ref{sample}). The biggest and most
complete IDV survey so far, the 5~GHz MASIV survey
\citep{Lovell_etal03}, started at the VLA in 2002; the first results
reported by \citeauthor{Lovell_etal03} suggest that 85 of 710
compact flat-spectrum sources are IDVs. The MASIV data, however, are
not yet fully published. We have labeled a source in our sample as an
IDV if there is a published statistically significant detection of flux
density variations on a time scale of less that 3 days (72 hours).
However, we are not able to identify all of the potential IDV sources
in our sample consistently, because some sources are not (yet) listed
in any of the published IDV survey results, and also because intra-day
variability is a transient phenomenon, and not all sources were
monitored equally well.


     Figure~\ref{IDV} shows the distribution of the sub-mas compactness
index, $S_\mathrm{unres}/S_\mathrm{VLBA}$; the median value over all
observing epochs was taken for each source. The median VLBA core
dominance, $S_\mathrm{core}/S_\mathrm{VLBA}$, and the maximum brightness
temperature are also shown. The full and the MOJAVE sample are shown
separately, and we have separated IDVs with high modulation index
($m>0.02$ observed at least once) from ones with low modulation index
(measured values of modulation index always less than 0.02, or not
reported).

\begin{figure*}[t]
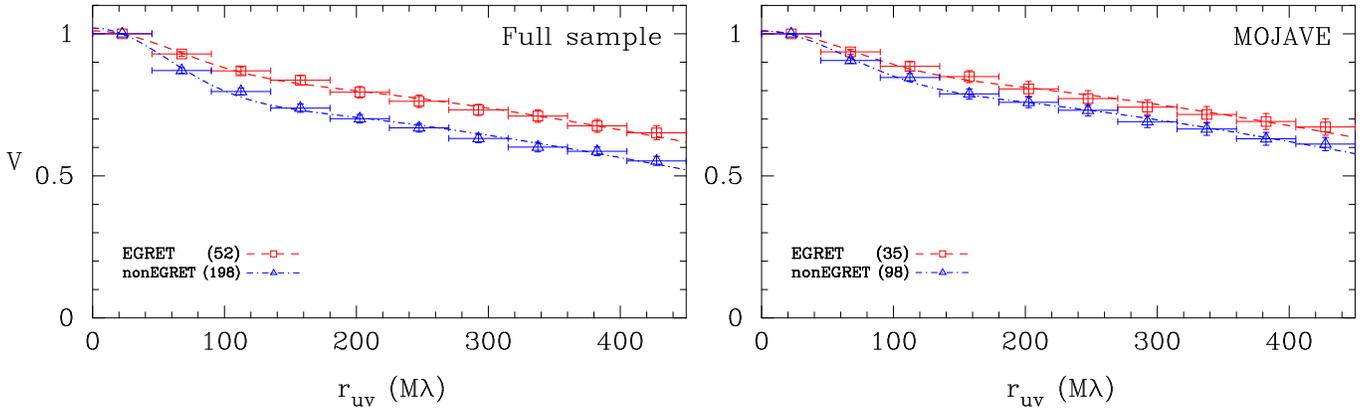

\resizebox{1.0\hsize}{!}{
\includegraphics[trim=0cm 0cm 0cm 0cm]{f13a_color.eps}
\includegraphics[trim=0.4cm 0cm 0cm 0cm,clip]{f13b_color.eps}
}
\caption{
The non-weighted mean of the fringe visibility versus projected
spacing. The distribution is normalized to 1.0 at 25 million
wavelengths. The visibility data at the epoch with the maximum
correlated flux density $S_\mathrm{unres}$ is used for each source. The
sample is divided into ``gamma-ray bright'' (EGRET detections) and
``gamma-ray weak'' (EGRET non-detections) sources (see
Table~\ref{sample} for details). The lines represent the best fitting
two-component models. See model parameters in Table~\ref{radplot-model}.
}
\label{relative_radplot_EGRET}
\end{figure*}

     We find that IDV sources have more compact and more core dominant
structure on sub-milliarcsecond scales (Table~\ref{IDV-table}) than
non-IDV sources. IDVs with a higher amplitude of intra-day variation
tend to have a higher unresolved flux density, $S_\mathrm{unres}$. The
results for core dominance are in agreement with previous findings
\citep[e.g.,][]{WQ93,Ojha_etal2004}.  A K-S test yields a probability
of less than 1\%\ for both the full and the MOJAVE sample that the
sub-mas compactness (Figure~\ref{IDV}) has the same parent distribution
for IDV and non-IDV sources.

     One might expect IDV behavior in almost all the sources with high
visibility amplitude at long VLBI spacings. However, this was not
observed by IDV surveys, perhaps because of the intermittent nature of
the IDV phenomenon.

     Some IDV observations have suggested apparent brightness
temperatures up to $10^{15}$\,K if they are due to interstellar
scintillations, and up to $10^{21}$\,K if they are intrinsic
\citep[e.g.,][]{KC_etal97}. More recently, IDV observations of
\cite{Lovell_etal03} have shown typical brightness temperature
values of the order of $10^{12}$\,K, consistent with our results
(Figure~\ref{IDV}). However, as seen from
equation~(\ref{Tb_eqn_notheta}), the highest brightness temperature
which we can reliably discern is of the order of $10^{13}$\,K, so we
are not able to comment on the evidence for the extremely high
brightness temperatures and the corresponding high Lorentz factors
inferred for some IDV.

\subsection{Gamma-ray Sources}

     The third catalog of high energy gamma-ray sources detected by the
EGRET telescope of the Compton Gamma Ray Observatory
\citep{Hartman_etal99} includes 66 high-confidence identifications of
blazars \citep{Mattox_etal01,SRM03,SRM04}. While the gamma-ray
sources were identified with flat-spectrum extragalactic radio sources
($\alpha>-0.5$, not all flat-spectrum sources have
been detected as gamma-ray sources. This is not necessarily indicative
of a bi-modality in the gamma-ray loudness distribution of
extragalactic radio sources (such as that found at radio wavelengths),
since the sensitivity level of EGRET was such that many sources were
only detected in their flaring state. With the next generation of
gamma-ray telescopes, such as GLAST \citep{GLAST}, the sensitivity may
be sufficient to actually separate the classes of gamma-ray loud and
gamma-ray quiet objects and to define the relationship between radio
and gamma-ray emission of the sources properly.

     For the purpose of our test we have grouped the ``highly
probable'' and ``probable'' EGRET identifications (Table~\ref{sample})
together, which yields 52 ``EGRET detections'' out of 250 objects for
the full and 35 out of 133 for the MOJAVE sample.

     In \citetalias{Kellermann_etal98} we did not find any clear
differences in the sub-milliarcsecond scale structure between the EGRET
(20\% of our radio sample) and non-EGRET sources. However, we find here
that the sub-mas compactness, $S_\mathrm{unres}/S_\mathrm{VLBA}$, for
EGRET detections is, on average, greater than for the EGRET
non-detections. This can be seen in Figure~\ref{relative_radplot_EGRET}
which shows the mean visibility function amplitudes versus projected
spacing for EGRET detected and non-detected sources, for the full and
MOJAVE samples. For both samples, the EGRET detected sources have, on
average, a higher contribution of compact VLBA structure (see
Table~\ref{radplot-model} for the parameters of the two-component fit).
This comparison is valid because the EGRET detected and non-detected
blazars in our sample have indistinguishable redshift distributions.
This result still holds if we exclude the GPS and steep spectrum
sources, which are generally gamma-ray weak \citep[see,
e.g.,][]{Mattox_etal97}. The difference is more pronounced for the full
sample, which is not selected on the basis of the VLBI flux density.
The difference for the MOJAVE sample is small, but remains significant.
These results suggest that a connection may exist between gamma-ray and
beamed radio emission from extragalactic sources on sub-milliarcsecond
scales, as has already being argued by others
\citep[e.g.,][]{Jorstad_etal01}.

\section{Summary}
\label{summary}

     We have analyzed visibility function data of 250 extragalactic
radio sources, obtained with the VLBA at 15 GHz. Almost all of the
radio sources in our sample have unresolved radio emission brighter
than 0.1~Jy on the longest VLBA baselines. For 171 objects, more than
half of the flux density comes from unresolved features. We have
compiled a list of 163 sources with unresolved structure stronger than
0.5~Jy, which will form a target list of special interest for planned
space VLBI observations such as RadioAstron, VSOP--2, and ARISE.

     A few of the sources have an overall radio structure which is only
slightly resolved at the longest spacings. Their total angular size is
less than about 0.05~mas, at least in one dimension, at some epochs.
However, even though most sources in our full sample are extended
overall, there are 158 sources in which the VLBA core component appears
unresolved, usually smaller than 0.05~mas, again in one direction, at
least at one epoch. For 19 of these, the core was unresolved at all
epochs.

     The distribution of the brightness temperature of the cores peaks
at $10^{12}$\,K and extends up to $5\,\times\,10^{13}$\,K; this is
close to the limit set by the dimensions of the VLBA. However, for many
sources we only measure a lower limit to the brightness temperature.
There is evidence that the observed brightness temperatures can be
explained as the result of Doppler boosting, but transient phenomena,
coherent emission, or synchrotron emission by relativistic protons may
also be important.

     On sub-milliarcsecond scales, active galaxies are on average
larger and less core dominated than quasars, which is consistent with
unification models in which the latter are viewed at smaller angles to
the line of sight. Additionally, the weak-lined objects classified as
BL Lacs tend to be smaller than the broad-lined quasars in our sample.
IDV sources show a higher compactness and core dominance on sub-mas
scales than non-IDV ones. IDVs with a higher amplitude of intra-day
variation tend to have a higher flux density in an unresolved
component. The most variable sources tend to have the most compact
structure. EGRET-detected radio sources show a higher degree of sub-mas
compactness than non-EGRET sources, supporting emission models which
relate the radio and gamma-ray emission, such as inverse Compton
scattering \citep[see, e.g.,][]{BloomMarscher96}.

\acknowledgments
     The National Radio Astronomy Observatory is a facility of the
National Science Foundation operated under cooperative agreement by
Associated Universities, Inc. The University of Michigan Radio
Astronomy Observatory has been supported by the University of Michigan
Department of Astronomy and the National Science Foundation. Part of
this work was done by MLL and DCH during their Jansky Postdoctoral
Fellowships at the National Radio Astronomy Observatory. This work was
supported partially by the NASA JURRISS program (W--19611), NSF
(0406923--AST), and by the Russian Foundation for Basic Research
(02--02--16305, 05--02--17377). MK was supported through a stipend from
the International Max Planck Research School for Radio and Infrared
Astronomy at the University of Bonn. This research has made use of the
NASA/IPAC Extragalactic Database (NED) which is operated by the Jet
Propulsion Laboratory, California Institute of Technology, under
contract with the National Aeronautics and Space Administration. We
thank the referee, Ski Antonucci, for many helpful comments, which have
led us in particular to clarify the relationship between source
classifications and our statistical results.



\clearpage
\LongTables

\begin{deluxetable}{llccllccclc}
\tablecolumns{11}
\tabletypesize{\scriptsize}
\tablewidth{0pt}
\tablecaption{
General Source Properties
\label{sample}}
\tablehead{
\colhead{IAU}  & \colhead {}      & \colhead {R.A.}     & \colhead {Dec.}     & \colhead{Opt.}    & \colhead{}    & \colhead{Radio}    & \colhead{EGRET} & \colhead{MOJAVE} & \colhead{VSOP}   & \colhead{IDV} \\
\colhead{Name} & \colhead {Alias} & \colhead {(J2000.0)}& \colhead {(J2000.0)}& \colhead{Class}   & \colhead{$z$} & \colhead{Spectrum} & \colhead{ID}    & \colhead{Member} & \colhead{Sample} & \colhead{Ref.}\\
\colhead{(1)}  & \colhead{(2)}    & \colhead{(3)}       & \colhead{(4)}       & \colhead{(5)}     & \colhead{(6)} & \colhead{(7)}      & \colhead{(8)}   & \colhead{(9)}    & \colhead{(10)}   & \colhead{(11)}
}
\startdata
0003$-$066 &  NRAO 5    & 00 06 13.8929 & $-$06 23 35.3353 & B\tablenotemark{b} & 0.347  & Flat   & \n & Y & A     & \n   \\
0007$+$106 &  III Zw 2  & 00 10 31.0059 & $+$10 58 29.5041 & G                  & 0.089  & Flat   & \n & Y & \n    & \n   \\
0014$+$813 &            & 00 17 08.4750 & $+$81 35 08.1360 & Q                  & 3.387  & Flat   & \n &\n & \n    & \n   \\
0016$+$731 &            & 00 19 45.7864 & $+$73 27 30.0175 & Q                  & 1.781  & Flat   & \n & Y & A, PR & \n   \\
0026$+$346 &            & 00 29 14.2425 & $+$34 56 32.2466 & G                  & 0.517  & Flat   & \n &\n & C     & \n   \\
0035$+$413 &            & 00 38 24.8436 & $+$41 37 06.0006 & Q                  & 1.353  & Flat   & \n &\n & C     & \n   \\
0039$+$230 &            & 00 42 04.5451 & $+$23 20 01.0610 & U                  & \n     & Peaked & \n &\n & A     & \n   \\
0048$-$097 &            & 00 50 41.3174 & $-$09 29 05.2102 & B                  & \n     & Flat   & \n & Y & \n    & \n   \\
\enddata
\end{deluxetable}
					  

\begin{deluxetable}{llrrrcrrrrrc}
\tablecolumns{12}
\tabletypesize{\scriptsize}		  
\tablewidth{0pt}			  
\tablecaption{Source parameters for an epoch of observations
when the flux density from the most compact component was maximum
\label{data_table}
}
\tablehead{				  
\colhead{Source}  & 
\colhead{Epoch}& \colhead{$S_\mathrm{tot}$} & \colhead{$S_\mathrm{VLBA}$} & \colhead{$S_\mathrm{unres}$} &
\colhead{$\frac{S_\mathrm{VLBA}}{S_\mathrm{tot}}$}  & \colhead{$S_\mathrm{core}$} & \colhead{$\theta_\mathrm{maj}$} & \colhead{$\theta_\mathrm{min}$} &
\colhead{$T_\mathrm{b}$} & $\mathrm{PA}_\mathrm{jet}$ & \colhead{\# of Epochs}\\
\colhead{(1)}  & 
\colhead{(2)}  & \colhead{(3)} & \colhead{(4)} & \colhead{(5)} &
\colhead{(6)}  & \colhead{(7)} & \colhead{(8)} & \colhead{(9)} &
\colhead{(10)} & \colhead{(11)} & \colhead{(12)}
}    	   				  
\startdata
0003$-$066                  & 2000/01/11 &  2.51 &  2.28 &  1.97 & 0.88 &  1.85 &    0.50 & $<$0.06 & $>$4.60e+11 &  $-$79 &  7 /  1 \\
0007$+$106                  & 1998/10/30 &  1.21 &  1.28 &  1.11 & 1.01 &  1.28 &    0.15 &    0.06 &    9.05e+11 &     \n &  6 /  4 \\
0014$+$813                  & 1999/01/02 &  0.50 &  0.47 &  0.24 & 0.94 &  0.37 &    0.54 &    0.14 &    1.20e+11 &    167 &  1 /  0 \\
0016$+$731                  & 2003/08/28 &  1.11 &  1.06 &  0.93 & 0.85 &  1.02 &    0.13 &    0.09 &    1.39e+12 &    130 &  6 /  2 \\
0026$+$346                  & 1995/04/07 &  0.70 &  0.65 &  0.21 & 0.66 &    \n &      \n &      \n &          \n &     55 &  7 /\n  \\
0035$+$413                  & 1998/03/19 &  0.50 &  0.47 &  0.29 & 0.93 &  0.38 &    0.43 &    0.18 &    6.16e+10 &    101 &  6 /  3 \\
0039$+$230                  & 1998/12/05 &  0.42 &  0.45 &  0.13 & 1.07 &  0.16 &    0.31 &    0.22 & $>$1.34e+10 & $-$164 &  1 /  0 \\
0048$-$097                  & 1996/10/27 &  2.09 &  1.60 &  1.60 & 0.85 &  1.61 &    0.08 & $<$0.05 & $>$2.37e+12 &  $-$12 &  5 /  2 \\
\enddata

\tablecomments{
The Table lists, for each source, parameters at the epoch for which its
unresolved flux density $S_\mathrm{unres}$ was greatest. The full
version of this table with data for all the epochs observed is
available in electronic form only.
Column designation:
(1)  IAU source name;
(2)  epoch of observations (YYYY/MM/DD);
(3)  filled-aperture (singe antenna) flux density interpolated to the epoch from column (2) (Jy);
(4)  total CLEAN flux density of map $S_\mathrm{VLBA}$ (Jy);
(5)  flux density from the most compact component of map $S_\mathrm{unres}$ (Jy);
(6)  as an exception, we present here {\it the median} value of $S_\mathrm{VLBA}/S_\mathrm{tot}$
     over all available epochs ($S_\mathrm{VLBA}/S_\mathrm{tot}$
     is only slightly variable if it is close to unity, which is the case for the most
     sources);
(7)  flux density, $S_\mathrm{core}$, for a VLBA core component (Jy);
(8)  FWHM major axis of the VLBA core component $\theta_\mathrm{maj}$ (mas);
(9)  FWHM minor axis of the VLBA core component $\theta_\mathrm{min}$ (mas);
(10) brightness temperature of a VLBA core component (K);
(11) position angle of the jet $\mathrm{PA}_\mathrm{jet}$ (degrees) defined as the
     median of the position angle of the second component location w.r.t.\
    the core position in our model from the multi-epoch observations;
(12) number of epochs of observations / number of epochs when the source core is
     unresolved at least in one direction.
}
\tablenotetext{a}{More than two components were used to model the structure.}

\end{deluxetable}

\begin{deluxetable}{lrllll}
\tabletypesize{\footnotesize}
\tablecaption{Results of two component modeling of mean
fringe visibility versus projected spacing dependence at 15 GHz
\label{radplot-model}
}
\tablewidth{0pt}
\tablehead{
\colhead{Sample} & \colhead{\# of sources} & \colhead{${\cal V}_1$} & \colhead{$\theta_1$} & \colhead{${\cal V}_2$} & \colhead{$\theta_2$} \\
\colhead{(1)} & \colhead{(2)} & \colhead{(3)} & \colhead{(4)} & \colhead{(5)} & \colhead{(6)} 
}
\startdata
MOJAVE sample: BL Lacs          &  22 & 0.16 & 1.353 & 0.85 & 0.105 \\
MOJAVE sampel: Quasars          &  94 & 0.16 & 1.278 & 0.85 & 0.144 \\
MOJAVE sample: Active galaxies  &   8 & 0.46 & 1.314 & 0.57 & 0.216 \\
MOJAVE sample: All              & 133 & 0.18 & 1.296 & 0.83 & 0.144 \\
\tableline	              
Full sample: EGRET       &  52 & 0.16 & 1.260 & 0.85 & 0.138 \\
Full sample: non-EGRET   & 198 & 0.26 & 1.394 & 0.76 & 0.151 \\
MOJAVE sample: EGRET     &  35 & 0.15 & 1.211 & 0.86 & 0.135 \\
MOJAVE sample: non-EGRET &  98 & 0.20 & 1.260 & 0.81 & 0.143 \\
\tableline	              
VSOP sample                  & 116 & 0.25 & 1.314 & 0.76 & 0.147 \\
VSOP Pearson-Readhead sample &  26 & 0.23 & 1.484 & 0.79 & 0.148 \\
\enddata
\tablecomments{
Column designation:
(1) Sample type;
(2) number of sources;
(3) peak value at zero spacing of the first Gaussian component;
(4) angular size of the first Gaussian component (mas);
(5) and (6)---the same for the second Gaussian component.
}
\end{deluxetable}

\begin{deluxetable}{lrccc}
\tabletypesize{\footnotesize}
\tablecaption{Parameters of IDV selected sub-samples
\label{IDV-table}
}
\tablewidth{0pt}
\tablehead{
\colhead{Sample} & \colhead{\# of Sources} & \colhead{$S_\mathrm{unres}/S_\mathrm{VLBA}$} & \colhead{$S_\mathrm{core}/S_\mathrm{VLBA}$} & \colhead{$\log(T_\mathrm{b,max})$} \\
\colhead{(1)} & \colhead{(2)} & \colhead{(3)} & \colhead{(4)} & \colhead{(5)}
}
\startdata
Full sample: high $m$ IDV   &  27 & 0.70$\pm$0.03 & 0.83$\pm$0.02 & 12.4$\pm$0.1 \\
Full sample: low $m$ IDV    &  16 & 0.62$\pm$0.04 & 0.73$\pm$0.03 & 12.2$\pm$0.2 \\
Full sample: non IDV        & 207 & 0.55$\pm$0.02 & 0.74$\pm$0.01 & 12.0$\pm$0.1 \\
MOJAVE sample: high $m$ IDV &  17 & 0.74$\pm$0.03 & 0.85$\pm$0.03 & 12.5$\pm$0.1 \\
MOJAVE sample: low $m$ IDV  &   8 & 0.68$\pm$0.03 & 0.77$\pm$0.04 & 12.5$\pm$0.2 \\
MOJAVE sample: non IDV      & 108 & 0.59$\pm$0.02 & 0.74$\pm$0.02 & 12.4$\pm$0.1 \\
\enddata
\tablecomments{
Column designation:
(1) Sample type;
(2) number of sources;
(3) average $S_\mathrm{unres}/S_\mathrm{VLBA}$ and its dispersion;
(4) average $S_\mathrm{core}/S_\mathrm{VLBA}$ and its dispersion,
(5) median value of $\log(T_\mathrm{b,max})$ and dispersion of the average value.
The 1$\sigma$ errors of the average values are calculated
assuming a normal distribution of these values.
}
\end{deluxetable}


\begin{thebibliography}{}
\bibitem[Aller et al.(1985)]{Aller_etal85}
Aller, H.~D., Aller, M.~F., Latimer, G.~E., \& Hodge, P.~E.\
1985, \apjs, 59, 513

\bibitem[Aller et al.(1992)Aller, Aller, \& Hughes]{AAH92}
Aller, M.~F., Aller, H.~D., \& Hughes, P.~A.\
1992, \apj, 399, 16

\bibitem[Aller et al.(2003)Aller, Aller, \& Hughes]{AAH03}
Aller, M.~F., Aller, H.~D., \& Hughes, P.~A.\
2003, in ASP Conf.\ Ser.\ 300, Radio Astronomy at the Fringe, ed.\
J.~A.\ Zensus, M.~H.\ Cohen, \& E.\ Ros (San Francisco: ASP), 159

\bibitem[Angel \& Stockman(1980)]{AS80}
Angel, J.~R.~P., \& Stockman, H.~S.\ 1980, \araa, 18, 321

\bibitem[Antonucci(1993)]{Antonucci93}
Antonucci, R.\
1993, \araa, 31, 473

\bibitem[Antonucci et al.(1987)]{Antonucci_etal87}
Antonucci, R.~R.~J., Hickson, P., Miller, J.~S., \& Olszewski, E.~W.\
1987, \aj, 93, 785 


\bibitem[Bertero \& De Mol(1996)]{BerteroDeMol96}
Bertero, M., \& De Mol, C.\
1996, in Progress in Optics, v.~XXXVI,
ed. E. Wolf (Elsevier: Amsterdam), 129

\bibitem[Bertsch(1998)]{Bertsch98}
Bertsch, D.\
1998, IAU Circ., 6807, 2

\bibitem[Best et al.(2003)]{Best_etal2003}
Best, P.~N., et al.\
2003, \mnras, 346, 1021

\bibitem[Bignall et al.(2002)]{Bignall_etal02}
Bignall, H.~E., et al.\
2002, PASA, 19, 29

\bibitem[Biretta et al.(1985)Biretta, Schneider, \& Gunn]{Biretta_etal85}
Biretta, J.~A., Schneider, D.~P., \& Gunn, J.~E.\
1985, \aj, 90, 2508

\bibitem[Bloom \& Marscher(1996)]{BloomMarscher96}
Bloom, S.~D., \& Marscher, A.~P.\
1996, \apj, 461, 657 


\bibitem[Carilli et al.(1998)]{Carilli_etal98}
Carilli, C.~L., et al.\
1998, \apj, 494, 175

\bibitem[Cohen et al.(1971)]{Cohen_etal71}
Cohen, M.~H., Cannon, W., Purcell, G.~H., Shaffer, D.~B.,
Broderick, J.~J., Kellermann, K.~I., \& Jauncey, D.~L.\
1971, \apj, 170, 207 

\bibitem[Cohen et al.(1975)]{Cohen_etal75}
Cohen, M. H., et al.\
1975, \apj, 201, 249


\bibitem[Dennett-Thorpe \& de Bruyn(2002)]{DB02}
Dennett-Thorpe, J., \& de Bruyn, A.~G.\
2002, Nature, 415, 57

\bibitem[Eracleous \& Halpern(2004)]{EracleousHalpern2004}
Eracleous, M., \& Halpern, J.~P.\
2004, \apjs, 150, 181


\bibitem[Frey et al.(2000)]{Frey_etal00}
Frey, S., Gurvits, L.~I., Altschuler, D.~R., Davis, M.~M., Perillat P.,
Salter, C.~J., Aller, H.~D., Aller, M.F., \& Hirabayashi H.\
2000, \pasj, 52, 975

\bibitem[Gehrels \& Michelson(1999)]{GLAST}
Gehrels, N., \& Michelson, P.\
1999, Astroparticle Physics, 11, 277

\bibitem[Greisen(1988)]{aips}
Greisen, E.~W.\
1988, in Acquisition, Processing and Archiving of Astronomical Images,
ed.\ G.~Longo, \& G.~Sedmak
(Napoli: Osservatorio Astronomico di Capodimonte), 125

\bibitem[Greve et al.(2002)]{Greve_etal02}
Greve, A., et al.\
2002, \aap, 390, L19


\bibitem[Hartman et al.(1999)]{Hartman_etal99}
Hartman, R.~C., et al.\
1999, \apjs, 123, 79

\bibitem[Heidt et al.(2004)]{Heidt_etal2004}
Heidt, J., et al.\
2004, \aap, 418, 813

\bibitem[Hirabayashi et al.(1998)]{Hirabayashi_etal98}
Hirabayashi, H., et al.\
1998, Science, 281, 1825

\bibitem[Hirabayashi et al.(2000)]{Hirabayashi_etal00}
Hirabayashi, H., et al.\
2000, \pasj, 52, 997

\bibitem[Hirabayashi et al.(2004)]{VSOP2}
Hirabayashi, H., Murata, Y., Edwards, P.~G., Asaki, Y.,
Mochizuki, N., Inoue, M., Umemoto, T., Kameno, S., \& Kono, Y.\
2004, in Proceedings of the 7th European VLBI Network Symposium,
ed.\ R.~Bachiller, F.~Colomer, J.-F.~Desmurs, P.~de~Vicente
(Observatorio Astronomico Nacional), 285; 
astro-ph/0501020

\bibitem[H\"ogbom(1974)]{Hogbom74}
H\"ogbom, J.~A.\
1974, \aaps, 15, 417

\bibitem[Homan et al.(2002)]{Homan_etal02}
Homan, D.~C., Ojha, R., Wardle, J.~F.~C., Roberts, D.~H.,
Aller, M.~F., Aller, H.~D., \& Hughes, P.~A.\
2002, \apj, 568, 99

\bibitem[Horiuchi et al.(2004)]{Horiuchi_etal2004}
Horiuchi, S., et al.\
2004, \apj, 616, 110

\bibitem[Jackson et al.(2002)]{Jackson_etal2002}
Jackson, C.~A., Wall, J.~V., Shaver, P.~A., Kellermann, K.~I.,
Hook, I.~M., \& Hawkins, M.~R.~S.\
2002, \aap, 386, 97

\bibitem[Jauncey \& Macquart(2001)]{JM01}
Jauncey, D.~L., \& Macquart, J.-P.\
2001, \aap, 370, L9 

\bibitem[Jorstad et al.(2001)]{Jorstad_etal01}
Jorstad, S. G., Marscher, A. P., Mattox, J. R., Wehrle, A. E.,
Bloom, S. D., \& Yurchenko, A. V.\
2001, \apjs, 134, 181

\bibitem[Kardashev(1997)]{RadioAstron}
Kardashev, N.~S.\
1997, Experimental Astron., 7, 329 

\bibitem[Kardashev(2000)]{Kardashev00}
Kardashev, N.~S.\
2000, ARep, 44, 719

\bibitem[Kataoka et al.(1999)]{Kataoka_etal99}
Kataoka, J., et al.\
1999, \apj, 514, 138

\bibitem[Kedziora-Chudczer et al.(1997)]{KC_etal97}
Kedziora-Chudczer, L., Jauncey, D.~L., Wieringa, M.~H.,
Walker, M.~A., Nicolson, G.~D.,  Reynolds, J.~E., \& Tzioumis, A.~K.\
1997, \apj, 490, L9

\bibitem[Kedziora-Chudczer et al.(2001)]{KC_etal01}
Kedziora-Chudczer, L., Jauncey, D.~L., Wieringa, M.~H.,
Tzioumis, A.~K., \& Reynolds, J.\
2001, \mnras, 325, 1411

\bibitem[Kellermann(2002)]{Kellermann02}
Kellermann, K.~I.\
2002, PASA, 19, 77

\bibitem[Kellermann(2003)]{Kellermann03}
Kellermann, K.~I.\
2003, in ASP Conf.\ Ser.\ 300, Radio Astronomy at the Fringe, ed.\
J.~A.\ Zensus, M.~H.\ Cohen, \& E.\ Ros (San Francisco: ASP), 185

\bibitem[Kellermann \& Pauliny-Toth(1969)]{Kellermann_Pauliny-Toth69}
Kellermann, K.~I., \& Pauliny-Toth, I.~I.~K.\
1969, \apj, 155, L31

\bibitem[Kellermann et al.(1973)]{Kellermann_etal73}
Kellermann, K.~I., Clark, B.~G., Cohen, M.~H., Shaffer, D.~B.,
Broderick, J.~J., \& Jauncey, D.~L.\
1973, \apj, 179, L141

\bibitem[Kellermann et al.(1998)]{Kellermann_etal98}
Kellermann, K.~I., Vermeulen, R.~C., Zensus, J.~A., \& Cohen, M.~H.\
1998, \aj, 115, 1295
\citepalias{Kellermann_etal98}

\bibitem[Kellermann et al.(2004)]{Kellermann_etal04}
Kellermann, K.~I., Lister, M.~L., Homan, D.~C., Vermeulen, R.~C.,
Cohen, M.~H., Ros, E., Kadler, M., Zensus, J.~A., \& Kovalev, Y.~Y.\
2004, \apj, 609, 539
\citepalias{Kellermann_etal04}


\bibitem[Kollgaard et al.(1992)]{Kollgaard_etal92}
Kollgaard, R.~I., Wardle, J.~F.~C., Roberts, D.~H., \& Gabuzda, D.~C.\
1992, \aj, 104, 1687


\bibitem[Kovalev(1998)]{Kovalev98}
Kovalev, Yu.~A.\
1998, Bull.~SAO, 44, 50

\bibitem[Kovalev et al.(1999)]{Kovalev_etal99}
Kovalev, Y.~Y., Nizhelsky, N.~A., Kovalev, Yu.~A., Berlin, A.~B.,
Zhekanis, G.~V., Mingaliev, M.~G., \& Bogdantsov, A.~V.\
1999, \aaps, 139, 545

\bibitem[Kovalev et al.(2000)]{Kovalev_etal2000}
Kovalev, Yu.~A., Kovalev, Y.~Y., \& Nizhelsky, N.~A.\
2000, \pasj, 52, 1027

\bibitem[Kovalev et al.(2002)]{KKNB02}
Kovalev, Y.~Y., Kovalev, Yu.~A., Nizhelsky, N.~A., \& Bogdantsov, A.~V.\
2002, PASA, 19, 83

\bibitem[Kraus et al.(2003)]{Kraus_etal03}
Kraus, A., et al.\
2003, \aap, 401, 161


\bibitem[Levy et al.(1989)]{Levy_etal89}
Levy, G.~S., et al.\
1989, \apj, 336, 1098

\bibitem[Lister(2001)]{Lister01}
Lister, M.~L.\ 2001, \apj, 562, 208

\bibitem[Lister et al.(2001)]{Lister_etal01}
Lister, M.~L., Tingay, S.~J. Murphy, D.~W., Piner, B.~G.,
Jones, D.~L. \& Preston, R.~A.\
2001a, \apj, 554, 948


\bibitem[Lister \& Homan(2005)]{MOJAVE-I}
Lister, M.~L.\, \& Homan, D.~C.\
2005, AJ, in press; astro-ph/0503152

\bibitem[Lobanov(2005)]{Lobanov2005}
Lobanov, A.~P.\
2005, \aap, submitted; astro-ph/0503225

\bibitem[Lobanov et al.(2000)]{Lobanov_etal00}
Lobanov, A.~P., Krichbaum, T.~P., Graham, D.~A., Witzel, A.,
Kraus, A., Zensus, J.~A., Britzen, S., Greve, A., \& Grewing, M.\
2000, \aap, 364, 391



\bibitem[Lovell et al.(2003)]{Lovell_etal03}
Lovell, J.~E.~J., Jauncey, D.~L., Bignall, H.~E.,
Kedziora-Chudczer, L., Macquart, J.-P., Rickett, B. J., \& Tzioumis, A.~K.\
2003, \aj, 126, 1699

\bibitem[Lovell et al.(2004)]{Lovell_etal2004}
Lovell, J.~E.~J., et al.\
2004, \apjs, 155, 27

\bibitem[Macomb et al.(1999)Macomb, Gehrels, \& Shrader]{Macomb_etal99}
Macomb, D.~J., Gehrels, N., \& Shrader, C.~R.\
1999, \apj, 513, 652

\bibitem[Macquart et al.(2000)]{Macquart_etal00}
Macquart, J.-P., Kedziora-Chudczer, L., Rayner, D.~P., \& Jauncey, D.~L.\
2000, \apj, 538, 623

\bibitem[Maltby \& Moffet(1962)]{MaltbyMoffet62}
Maltby, P., \& Moffet, A.~T.\
1962, \apjs, 7, 141

\bibitem[Marcha \& Browne(1995)]{MB95}
Marcha, M.~J.~M., \&  Browne, I.~W.~A.\
1995, \mnras, 275, 951 

\bibitem[Mattox et al.(1997)]{Mattox_etal97}
Mattox, J.~R., Schachter, J., Molnar, L., Hartman, R.~C., \& Patnaik, A.~R.\
1997, \apj, 481, 95

\bibitem[Mattox et al.(2001)Mattox, Hartman, \& Reimer]{Mattox_etal01}
Mattox, J.~R., Hartman, R.~C., \& Reimer, O.\
2001, \apjs, 135, 155





\bibitem[Moellenbrock et al.(1996)]{Moellenbrock_etal96}
Moellenbrock, G.~A., et al.\
1996, \aj, 111, 2174


\bibitem[Napier et al.(1994)]{VLBA}
Napier, P.~J., Bagri, D.~S., Clark, B.~G., Rogers, A.~E.~E., \& Romney, J.~D.\
1994, Proc.~IEEE, 82, 658


\bibitem[Ojha et al.(2004)]{Ojha_etal2004}
Ojha, R., Fey, A.~L., Jauncey, D.~L., Lovell, J.~E.~J., \& Johnston, K.~J.\
2004, \apj, 614, 607

\bibitem[Pearson(1999)]{Pearson99}
Pearson, T.~J.\
1999, in ASP Conf.~Ser.\ 180, Synthesis in Radio Astronomy II,
ed.\ G.~ B.~Taylor, C.~L.~Carilli, \& R.~A.~Perley
(San Francisco: ASP), 335

\bibitem[Pearson \& Redhead(1984)]{PearsonRedhead84}
Pearson, T.~J., \& Readhead, A.~C.~S.\
1984, \araa, 22, 97


\bibitem[Quirrenbach et al.(1992)]{Quirrenbach_etal92}
Quirrenbach, A., et al.\
1992, \aap, 258, 279

\bibitem[Quirrenbach et al.(2000)]{Quirrenbach_etal2000}
Quirrenbach, A., et al.\
2000, \aaps, 141, 221

\bibitem[Readhead(1994)]{Readhead94}
Readhead, A.~C.~S.\
1994, \apj, 426, 51


\bibitem[Rector \& Stocke(2001)]{RS2001}
Rector, T.~A., \& Stocke, J.~T.\
2001, \aj, 122, 565

\bibitem[Rickett et al.(2001)]{Rickett_etal01}
Rickett, B. J., Witzel, A., Kraus, A., Krichbaum, T.~P., \& Qian, S.~J.\
2001, \apj, 550, L11 

\bibitem[Rowson(1963)]{Rowson63}
Rowson, B.\
1963, \mnras, 125, 177 

\bibitem[Scarpa \& Falomo(1997)]{SF97}
Scarpa, R., \&  Falomo, R.\ 1997, \aap, 325, 109 

\bibitem[Scott et al.(2004)]{Scott_etal2004}
Scott, W.~K., et al.\
2004, \apjs, 155, 33

\bibitem[Shepherd(1997)]{difmap}
Shepherd,~M.~C.\
1997, in ASP Conf.\ Series.\ 125,
Astronomical Data Analysis Software and Systems~VI,
ed.\ G.~Hunt \& H.~E.~Payne (San Francisco: ASP), 77

\bibitem[Singal \& Gopal-Krishna(1985)]{SG85} Singal,
K.~A.~\& Gopal-Krishna 1985, \mnras, 215, 383

\bibitem[Small et al.(1997)Small, Sargent, \& Steidel]{Small_etal97}
Small, T.~A., Sargent, W.~L.~W., \& Steidel, C.~C.\
1997, \aj, 114, 2254


\bibitem[Sowards-Emmerd et al.(2003)Sowards-Emmerd, Romani, \& Michelson]{SRM03}
Sowards-Emmerd, D., Romani, R.~W., \& Michelson, P.~F.\
2003, \apj, 590, 109

\bibitem[Sowards-Emmerd et al.(2004)]{SRM04}
Sowards-Emmerd, D., Romani, R.~W., Michelson, P.~F., \& Ulvestad, J.~S.\
2004, \apj, 609, 564

\bibitem[Stickel et al.(1991)]{SFK91}
Stickel, M., Fried, J.~W., K\"uhr, H., Padovani, P., \& Urry, C.~M.\ 1991, \apj, 374, 431 

\bibitem[Stickel et al.(1994)Stickel, Meisenheimer, \& K\"uhr]{Stickel_etal94}
Stickel, M., Meisenheimer, K., \& K\"uhr, H.\
1994, \aaps, 105, 211


\bibitem[Stocke \& Rector(1997)]{StockeRector97}
Stocke, J.~T., \& Rector, T.~A.\
1997, \apj, 489, L17


\bibitem[Tingay et al.(2001)]{Tingay_etal01}
Tingay, S.~J., et al.\
2001, \apj, 549, L55

\bibitem[Tornikoski et al.(1999)]{Tornikoski_etal99}
Tornikoski, M., et al.\
1999, \aj, 118, 1161

\bibitem[Ulvestad(2000)]{ARISE}
Ulvestad, J.~S.\
2000, Adv.\ Space Research, 26, 735 

\bibitem[Urry \& Padovani(1995)]{UrryPadovani95}
Urry, C.~M., \& Padovani, P.\
1995, \pasj, 107, 803


\bibitem[Vermeulen et al.(1995)]{Vermeulen_etal95}
Vermeulen, R.~C., et al.\
1995, \apj, 452, L5 

\bibitem[Vermeulen et al.(2003)]{VRK03}
Vermeulen, R.~C., Ros, E., Kellermann, K.~I., Cohen, M.~H., Zensus,
J.~A., \& van Langevelde, H.~J.\
2003, \aap, 401, 113

\bibitem[V{\'e}ron-Cetty \& V{\'e}ron(2000)]{VCV00}
V{\'e}ron-Cetty, M.~P., \& V{\'e}ron, P.\ 2000, \aapr, 10, 81 

\bibitem[V{\'e}ron-Cetty \& V{\'e}ron(2003)]{VCV03}
V{\'e}ron-Cetty, M.-P., \& V{\' e}ron, P.\
2003, \aap, 412, 399 


\bibitem[Whitney et al.(1971)]{Whitney_etal71}
Whitney, A.~R., Shapiro, I.~I., Rogers, A.~E.~E., Robertson, D.~S.,
Knight, C.~A., Clark, T.~A., Goldstein, R.~M., Marandino, G.~E.,
\& Vandenberg, N.~R.\
1971, Science, 173, 225


\bibitem[Wills(1999)]{Wills99}
Wills, B.~J.\
1999, ASP Conf.~Ser.~162, Quasars and Cosmology,
ed.\ G.~Ferland, \& J.~Baldwin, (San Francisco: ASP), 101

\bibitem[Wills et al.(1983)]{Wills_etal83}
Wills, B.~J., et al.\
1983, \apj, 274, 62

\bibitem[Witzel \& Quirrenbach(1993)]{WQ93}
Witzel, A., \& Quirrenbach, A.\
1993, in Proceedings of the workshop `Propagation Effects in Space VLBI'
held in Leningrad, June 1990, ed.\ L.~I.~Gurvits
(Arecibo Observatory: NAIC), 33

\bibitem[Zensus et al.(2002)]{Zensus_etal02}
Zensus, A., Ros, E., Kellermann, K.~I., Cohen, M.~H., \& Vermeulen, R.~C.\
2002, \aj, 124, 662
\citepalias{Zensus_etal02}
\end{thebibliography}
\end{document}